# Mapping the Diffusion Tensor in Microstructured Perovskites

*Roberto Brenes,[1,2] Dane W. deQuilettes[1*] Richard Swartwout,[1] Abdullah Y. Alsalloum,[3] Osman M. Bakr,[3] Vladimir Bulović[1,2*]*

[1] Research Laboratory of Electronics, Massachusetts Institute of Technology, 77 Massachusetts Avenue, Cambridge, Massachusetts 02139, USA

[2] Department of Electrical Engineering and Computer Science, Massachusetts Institute of Technology, 77 Massachusetts Avenue, Cambridge, Massachusetts 02139, USA

[3] Division of Physical Sciences and Engineering, KAUST Catalysis Center (KCC), King Abdullah University of Science and Technology, Thuwal 23955-6900, Kingdom of Saudi Arabia

*Corresponding Authors: danedeq@mit.edu; bulovic@mit.edu

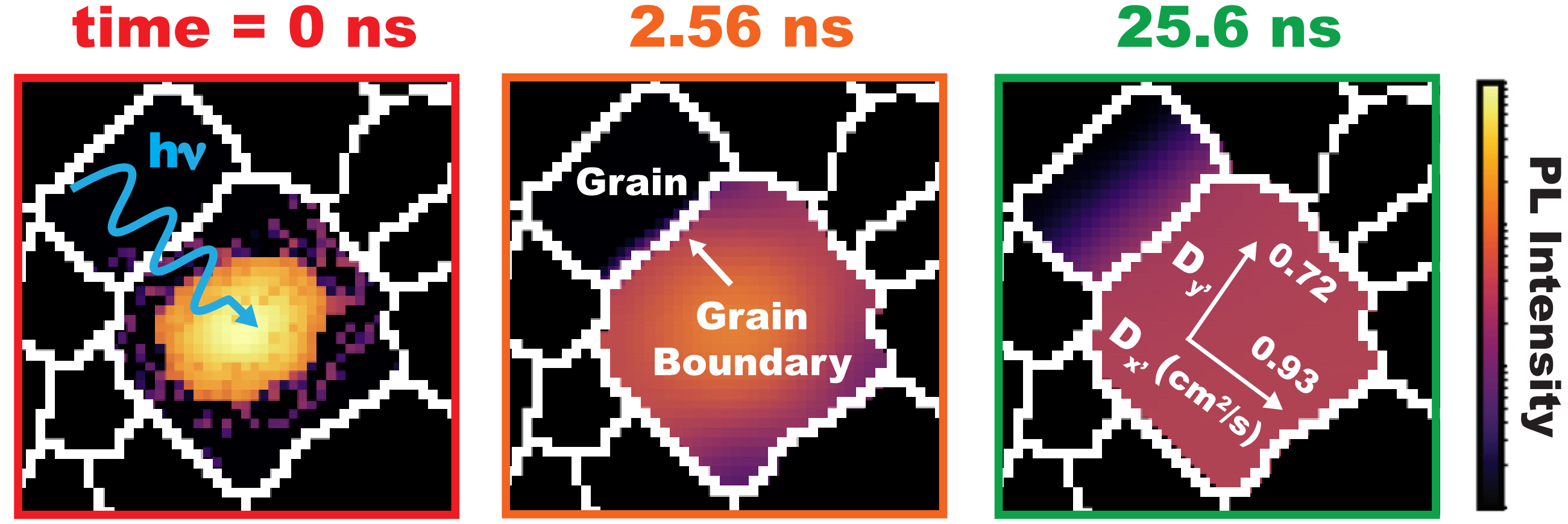

*Understanding energy transport in semiconductors is critical for design of electronic and optoelectronic devices. Semiconductor material properties such as charge carrier mobility or diffusion length are commonly measured in bulk crystals and determined using models that describe transport behavior in homogeneous media, where structural boundary effects are minimal. However, most emerging semiconductors exhibit nano and microscale heterogeneity. Therefore, experimental techniques with high spatial resolution paired with models that capture anisotropy and domain boundary behavior are needed. We develop a diffusion tensor-based framework to analyze experimental photoluminescence (PL) diffusion maps accounting for material nano and microstructure. Specifically, we quantify both carrier transport and recombination in single crystal and polycrystalline lead halide perovskites by globally fitting diffusion maps, with spatial, temporal, and PL intensity data. We reveal a 29% difference in principal diffusion coefficients and alignment between electronically coupled grains for* $CH_3NH_3PbI_3$ *polycrystalline films. This framework allows for understanding and optimizing anisotropic energy transport in heterogeneous materials.*

Energy transport is an important mechanism to understand for the design and performance evaluation of photovoltaics, display technologies, transistors, optical sensors and sources.[1,2] Crucially, many semiconductors, ranging from inorganic,[3–6] organic[7–10] and lead halide perovskites[11–15] have been shown to exhibit various forms of electrical energy transport anisotropy, which include charge carriers as well as neutral excitons[16]. Understanding a material's diffusive anisotropy relative to crystallographic orientation could lead to improved photovoltaic devices and light-emitting diodes by preferential crystal growth and optimization.[17–19] Furthermore, heterogeneous interfaces can contribute to anisotropic energy transport,[13,15,20,21] a phenomenon which may be hard to differentiate from intrinsic anisotropy due to crystallographic orientation.[22,23] This knowledge gap can severely limit the performance of polycrystalline lateral devices where transport occurs across interfaces such as field effect transistors and back-contact photovoltaics.[24–27] So far, studies quantifying anisotropy have been carried out by analyzing the expansion of photogenerated carrier profiles along different axes after focused laser excitation.[7,11,13] Through this approach, the interplay between diffusive transport and morphological effects is lost and thus difficult to differentiate.[23] Using a partial differential equation (PDE) model with a diffusion tensor to characterize carrier diffusion in every direction and taking into account the specific material geometry would capture all of these effects. The advent of high performance computing services[28] and efficient parallel computation as well as multi-dimensional optical techniques with high sensitivity[13,29,30] has made diffusion tensor modeling of multi-dimensional transport data feasible.

In this work, we develop a diffusion tensor-based framework for modeling carrier transport in nano and microstructured materials where grain boundary and morphological effects might be present. Using a diffusion tensor and the material microstructure geometry, we can effectively differentiate the morphological and anisotropy effects to reproduce not only the spatial shape but the signal intensity of the experimental data with a high degree of accuracy. This indicates that our model not only captures carrier transport effects, but also recombination dynamics. Through this diffusion tensor, we also determine the principal diffusion directions in polycrystalline films and their relative misorientation between grains. Due to the generality of the framework, which is based on a PDE model, the method can be adapted for any arbitrary micro- or nanostructured material to evaluate whether anisotropic energy transport is present.[3,31] In addition, by coupling this framework with a time-resolved scanning photoluminescence imaging measurement, we can resolve nanoscale changes in the carrier distributions on the order of 2nm even if the initial carrier excitation profile is limited by the diffraction limit of the laser illumination spot size (~600nm)[32].

## RESULTS AND DISCUSSION

In order to model higher-order recombination, anisotropy and film structure effects, we use a non-linear PDE model similar to our previous work:[23]

$$\frac{\partial \mathrm{N}(\boldsymbol{u},t)}{\partial t} = \nabla \cdot (\boldsymbol{D}(\boldsymbol{u})\nabla \mathrm{N}(\boldsymbol{u},t)) - k_1 \mathrm{N}(\boldsymbol{u},\mathrm{t}) - P_{esc} k_2 \mathrm{N}^2(\boldsymbol{u},t) - k_3 \mathrm{N}^3(\boldsymbol{u},t) \quad (1)$$

where $\boldsymbol{D}$ is a 2x2 diffusion tensor; $\nabla$ is the gradient operator; $\mathrm{N}(\boldsymbol{u},t)$ is the carrier density as a function of position $\boldsymbol{u} = (x,y)$ in the measurement axes and time; $k_1$ is the non-radiative, first-order (monomolecular) recombination constant, $k_2$ is the second-order (bimolecular) recombination rate constant, $k_3$ is the third-order (Auger) recombination rate constant and $P_{esc}$ the average probability of escape for a photon emitted inside the film. $\boldsymbol{D}$ and $k_1$ are fitted parameters while $k_2$, $k_3$ and $P_{esc}$ were taken from literature. The value for $k_3$was not considered for the fits since it did not impact the fitted results at the carrier densities probed (see Section S1. and S9. for details). In this scenario, we are able to capture anisotropic diffusion through use of the 2x2 diffusion tensor for $\boldsymbol{D}$. Furthermore, we model diffusion as a two-dimensional process due to small film thicknesses (<1 $\mu$m). In this thickness limit, charge carriers rapidly distribute through the film thickness in ~ 1 ns after photoexcitation, which is comparable to the time resolution of the measurement.[33]

From above, $\boldsymbol{D}$ quantifies the amount of anisotropy in carrier diffusion and has the explicit form:[22]

$$\boldsymbol{D} = \begin{bmatrix} D_{xx} & D_{xy} \\ D_{yx} & D_{yy} \end{bmatrix} \tag{2}$$

Where the diagonal terms, $D_{xx}$ and $D_{yy}$, are the diffusivities in the *x* and *y* directions in the measurement coordinates, respectively; and the off-diagonal terms, $D_{xy}$ is the diffusivity along the *x* direction arising from a concentration in the *y* direction and $D_{yx}$ is the diffusivity along the *y* direction arising from a concentration in the *x* direction. Note that the tensor must have conjugate symmetry, therefore $D_{xy} = D_{yx}$ .[34]

If the off-diagonal values are not zero, the diffusion in the system is anisotropic, and the second-rank tensor $\boldsymbol{D}$ can be converted to its two principal axes by diagonalizing the tensor:

$$\boldsymbol{D'} = \begin{bmatrix} D_{x\prime} & 0 \\ 0 & D_{y\prime} \end{bmatrix} = R \cdot \boldsymbol{D} \cdot R^T \tag{3}$$

where *x'* and *y'* are the principal diffusion axes, $D_{x\prime}$ and $D_{y\prime}$ are the diffusivities in the *x'* and *y'* direction corresponding to the eigenvalues of $\boldsymbol{D}$, and *R* is the rotation matrix where its columns are composed of the eigenvectors of $\boldsymbol{D}$ (see SI for details).[22,34–36]

Figure 1 shows the general schematic for quantifying anisotropy and modeling the measured data with the PDE model in Eqs. (1) and (2). Spatial and time-resolved data that measures energy transport is used as the main input for the model. For accurately quantifying anisotropic transport, the data must at least capture two spatial dimensions, since the data is globally fit for all measured time-resolved pixels (i.e. 2500 total pixels for $MAPbBr_3$ and 902 total pixels for $MAPbI_3$) in the photoluminescence map simultaneously to identify any interdependence between spatial dimensions. Since microstructure can play a significant role in energy transport, we also identify and input features of the microstructure (ie. domain boundaries) of the film for the geometry of the PDE model.[15,20,21,23,30,37] The microstructure surrounding the region where the diffusion map is collected by defining domain boundaries using a watershed segmentation algorithm in MorphoLibJ[38]. For convenience, we use a photoluminescence (PL) map as the input, but maps and images from nanoscale techniques such as atomic force microscopy, scanning electron microscopy, and/or electron backscatter diffraction may be used as well.[39] The edges are then vectorized into a closed polygon that can be used as the PDE geometry. In our specific case, we only modeled adjacent grains that were deemed electronically coupled through widefield PL imaging to save on computational costs (see Fig. S1).[15] We note that we do not claim that these optically-resolved boundaries are necessarily grain boundaries, since this has been reported to not be the case.[39] As initial conditions for the simulation, the carrier density and profile are determined from the absorbed laser power and the initial frame of the diffusion data, therefore non-idealities in the laser profile are taken into account in the model (see SI for details).[30,40]

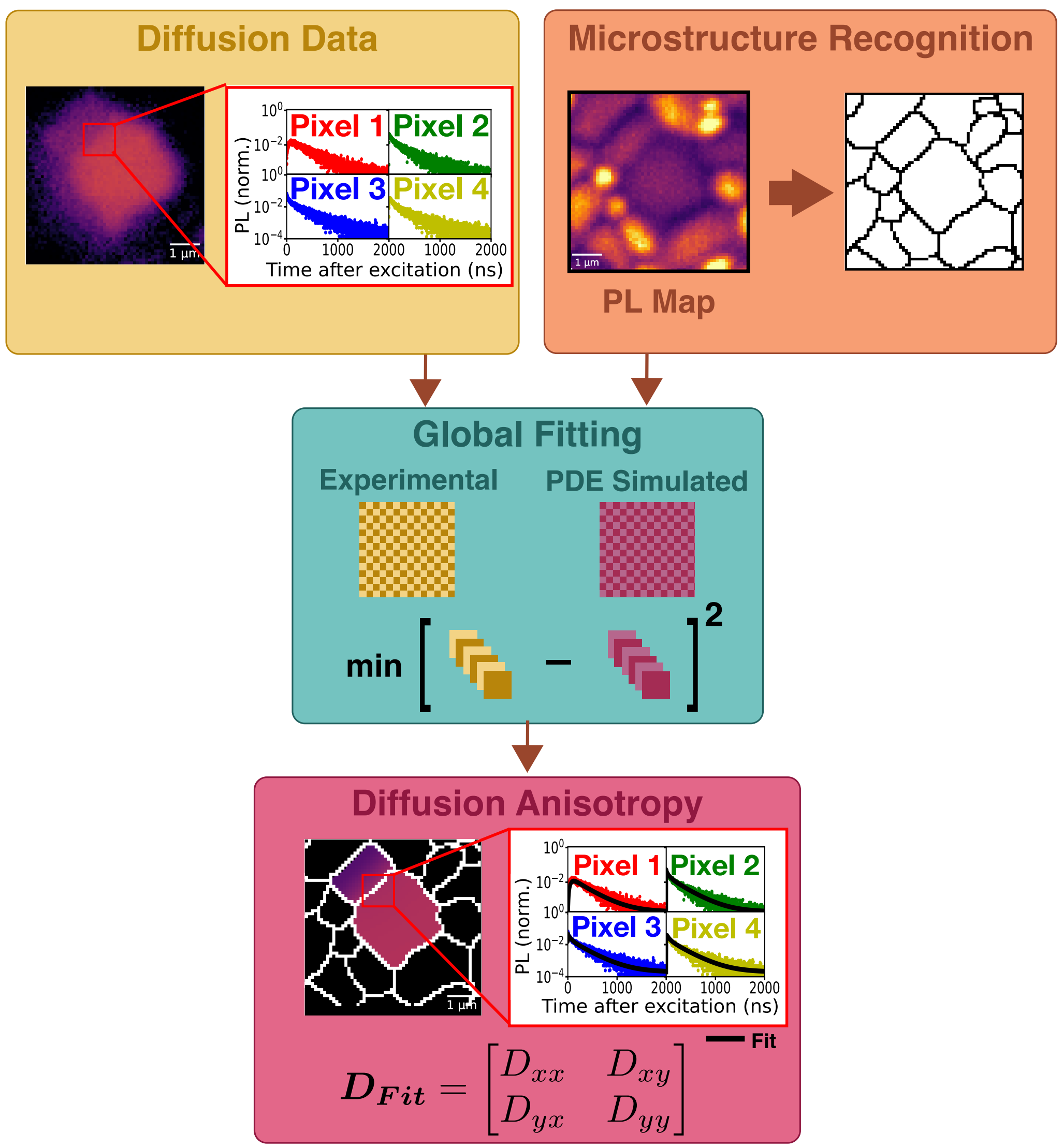

**Figure 1.** Schematic of the framework for modeling anisotropy in microstructured semiconducting films. The first input is the measured diffusion data, in this case through time-resolved photoluminescence (PL) microscopy. The selected area in the PL image shows that each pixel contains time-resolved PL data. The second input is the microstructure of the material, which was determined through PL confocal mapping and an edge-finding algorithm. The experimental data is then globally fit for all pixels with a non-linear least squares minimization. The simulations use a partial differential equation (PDE) model that utilizes the geometry determined from the microstructure. The model then outputs a fitted diffusion tensor and the simulated data with the fitted parameters. The simulation data has the same dimensionality as the input diffusion data.

After the base PDE model is determined, a non-linear least squares minimization is carried out between the experimental data and the simulation, while fitting $\boldsymbol{D}$ and $k_1$. All other parameters were kept constant using widely reported literature values (see SI for details)[41–46]. The minimization is carried out over all data points (both spatial and temporal) simultaneously, following the least-squares minimization problem:

$$min \sum_{t} \sum_{\boldsymbol{u}} \left( \frac{N(\boldsymbol{u},t)^2}{max(N(\boldsymbol{u},t)^2)} - \frac{PL_{exp}(\boldsymbol{u},t)}{max\left(PL_{exp}(\boldsymbol{u},t)\right)} \right)^2 \tag{4}$$

Where $PL_{exp}(\boldsymbol{u},t)$ is the spatially and temporally dependent experimental photoluminescence. Both the simulated and experimental data have been scaled by their corresponding maximum value to assign higher weighting to higher signals.

This ensures that both the diagonal and off-diagonal components of $\boldsymbol{D}$ are accurately determined with grain boundary effects taken into account. Without this scheme, the carrier profiles might appear to be affected solely by diffusion effects when in reality it might be due to boundary effects, as we have previously shown is the case with MSD profiles[23]. The model then outputs the optimized simulation, along with the fitted 2x2 diffusion tensor $\boldsymbol{D}$, the diffusion coefficients for adjacent grains and the monomolecular recombination constant, $k_1$.

In order to generate multi-dimensional diffusion data, we performed all measurements using a custom diffusion mapping setup equipped with a 2-axis scanning avalanche photodiode (APD) (see SI for further details). We first validate the framework described in Figure 1 by measuring an ultrathin (~880 nm, see Fig. S2) $CH_3NH_3PbBr_3$ ($MAPbBr_3$) single crystal (see SI for preparation details) to remove vertical diffusion and microstructure effects. Figure 2a and b show an optical transmission image of the $MAPbBr_3$ single crystal along with its cubic crystal structure (i.e. a=a=a - symmetric geometry) at room temperature. We specifically chose MAPbBr3 as we expect the diffusion to be isotropic for this material. [22,47,48] Figures 2b and c display two individual frames of the measured diffusion data in a 5 $\mu$m by 5 $\mu$m area while Figures 2e and f show the same frames in the fitted simulation. Figure 2g demonstrates representative individual pixel fits of the total 2500 time-resolved pixel fits for the regions denoted with colored "x"s in b-e, we report excellent agreement in both the shape and absolute intensity of the simulated and experimental data. The shape is primarily captured by the diffusion coefficient and the intensity is dependent on the diffusion coefficient, recombination rate constants, and local carrier densities. By globally fitting the entire map, we are able to constrain the model PDE very well as slight variations in each of these parameters lead to large changes in the fit error. Through global minimization of the least squares cost function, we report a $D_{xx} = 0.95 \pm 0.03$ $cm^2s^{-1}$ (see SI for details on uncertainty) and $D_{yy} = 0.93 \pm 0.03$ $cm^2s^{-1}$ with off-diagonal components of zero as well as $k_1 = 3.34 \pm 0.01 \times 10^6$ $s^{-1}$. These results and the off-zero diagonal indicate that the diffusion is isotropic in this material with no coupling between the $x$ and $y$ dimensions. We note that we confirmed through PL mapping that the measured area showed high spatial PL uniformity, so boundary effects should not be present (see Fig. S3). Indeed, the extracted diffusion coefficients agree with others reported in literature and are consistent with measured carrier mobilities of around 40 $cm^2V^{-1}s^{-1}$ through the Einstein relation $\mu \cdot kT/q = D$.[40,49–52]

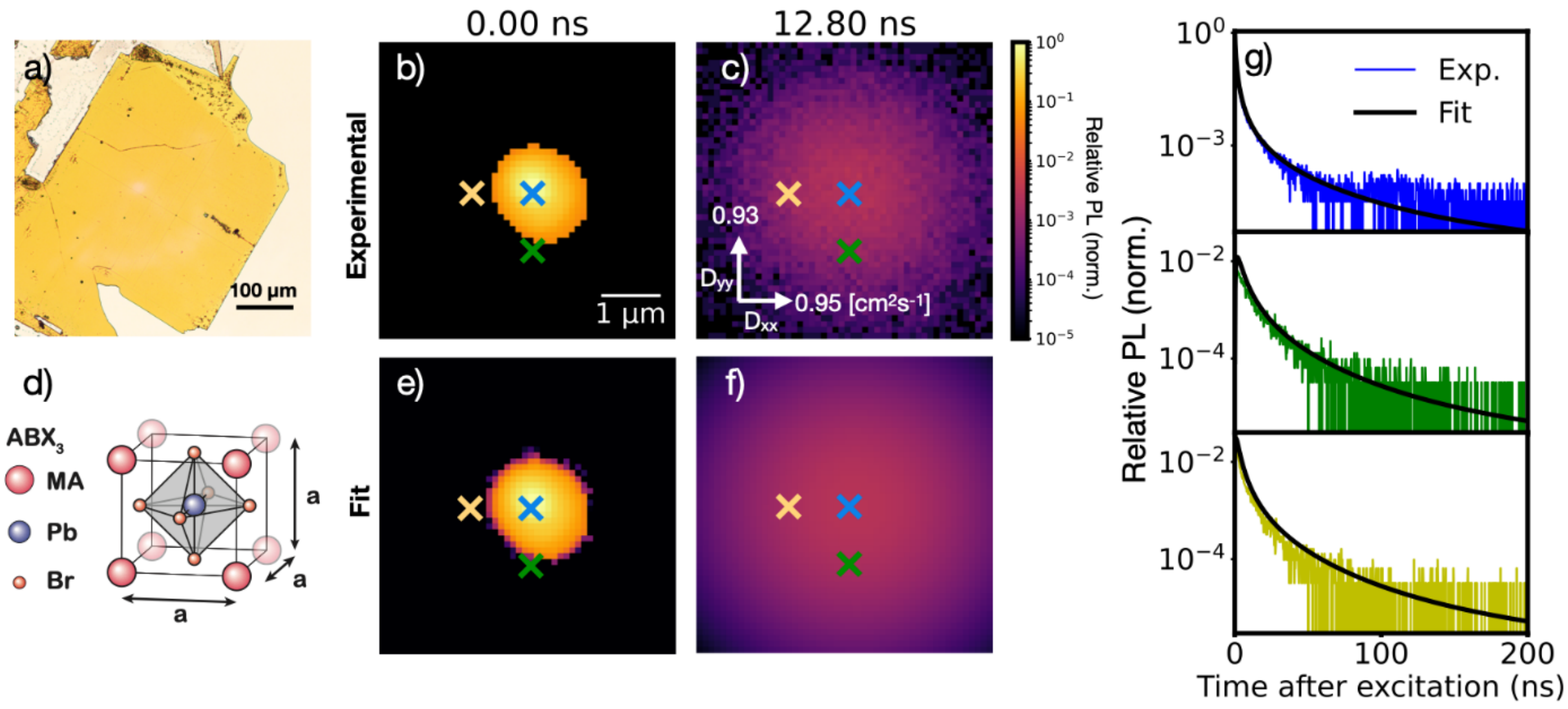

**Figure 2.** a) Optical transmission image of the ultrathin (~ 880 nm) $MAPbBr_3$ single crystal used for diffusion measurements b-c) Experimental two-dimensional photoluminescence (PL) profiles at b) 0 ns and c) 12.8 ns. The inset axes in c) show the fitted non-zero components of the diffusion tensor. The initial frame shows the laser excitation profile. d) Cubic crystal structure of $MAPbBr_3$ at room temperature, with lattice constant $a$=$a$=$a$ e-f) Simulated two-dimensional PL profiles corresponding to the same frames as b-c). b-c) and e-f) share the same scalebar. g) Time-resolved photoluminescence data and optimized simulation result corresponding to the sampled pixel locations of the colored "x"s in b-c, e-f). All images and traces are normalized relative to the maximum pixel value at t = 0 ns.

Figure 2 demonstrates that the framework described in Figure 1 is accurate when boundary conditions are not present. In order to extend the approach to a wider range of nano and microstructured materials, we next study a polycrystalline $CH_3NH_3PbI_3$ ($MAPbI_3$) thin-film (thickness ~ 330 nm). The film presents electronic coupling only with one adjacent grain (see Fig. S1), a selectivity that has been reported before.[13,15] The coupled grains are modeled by separate subdomains with different diffusion coefficients but the same monomolecular recombination rate constant (see SI for details). All other boundaries are considered reflective, as has been reported before.[13,21] For simplicity and minimizing the number of fitted parameters, only the illuminated grain was fitted with anisotropy, since we can capture the transport of carriers along multiple dimensions surrounding the excitation spot (see Fig. S1).

Figures 3b-d show selected frames of the measured two-dimensional diffusion profiles. The diffusion data exhibits fast spreading to the edges of the illuminated grain, within the first few nanoseconds (c.f. Supplementary Video 1). However, the adjacent electronically-coupled grain takes significantly longer (~20 ns) to start showing any detectable PL. Figures 3e-g shows the optimized simulation for the same frames as Figures 3b-d, with the grain boundaries added in white as a visual aid. The simulation shows the same behavior, where carriers diffuse quickly within the illuminated grain, but slower in the coupled grain (c.f. Supplementary Video 1). Figure 3h shows time-resolved PL traces for representative selected pixels highlighted with colored circles in both the experimental and simulated data out of the total 902 globally fitted time-resolved pixels. The simulations show excellent agreement with the experimental data, both in terms of intensity and shape of the TRPL traces. The early time dynamics show a rise in photoluminescence for pixels far away from the excitation spot, indicative of carrier transport (see Fig. S7 for data and fits on an expanded timescale)[53]. The fitted 2x2 diffusion tensor is:

$$\boldsymbol{D_{fit}} = \begin{bmatrix} 0.86 \pm 0.02 & 0.10 \pm 0.02 \\ 0.10 \pm 0.02 & 0.79 \pm 0.02 \end{bmatrix} \quad [\mathrm{cm^2 s^{-1}}] \tag{5}$$

The non-zero off-diagonal components indicate anisotropy and a misalignment between the principal axes ($x'$, $y'$) and the measurement axes ($x$, $y$). By diagonalizing Eq. (4) as seen on Eq. (3), we can determine the diffusion tensor $\boldsymbol{D_P}$ along the principal axes ($x'$, $y'$) (see SI for details):

$$\boldsymbol{D_P} = \begin{bmatrix} 0.93 \pm 0.04 & 0 \\ 0 & 0.72 \pm 0.04 \end{bmatrix} \quad [\mathrm{cm^2 s^{-1}}] \tag{6}$$

where the principal axes ($x'$, $y'$) are at an angle of $\theta$ = -34.4 ± 2.2° relative to the measurement axes ($x$, $y$). The inset axes on Figure 3c show the determined principal axes relative to the reference frame of the measurement. The adjacent electronically-coupled grain shows a much smaller fitted diffusion coefficient of $D$ = 0.11 ± 0.02 $cm^2s^{-1}$. The fitted diffusion coefficients for both the illuminated and non-illuminated grain fall in the range of reported values for $MAPbI_3$ thin-films between 0.004 - 3.3 $cm^2s^{-1}$.[13,21,30,54–61] This smaller "effective" diffusion coefficient in non-illuminated grains (see also Fig. S4) could be caused either by grain boundaries representing a carrier bottleneck as has been explored before[13], or due to a carrier dependent diffusion coefficient (see Fig. S5).[58]

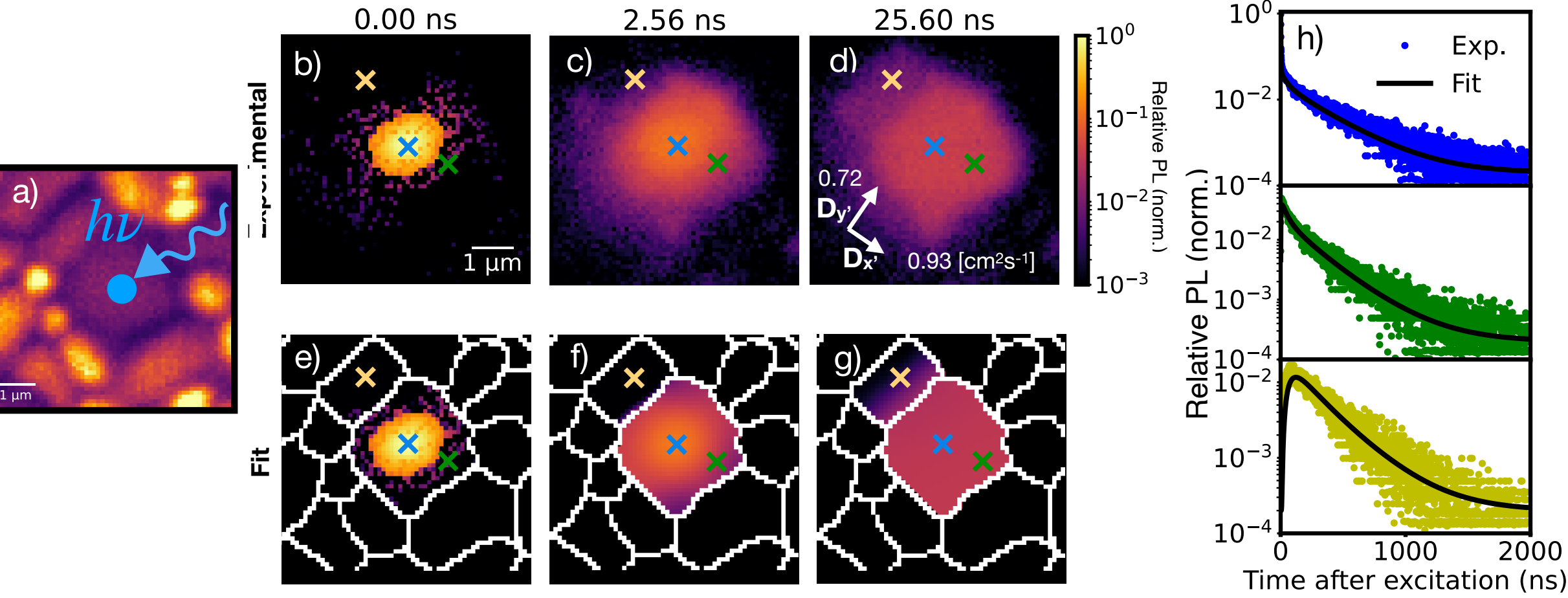

**Figure 3. a)** Photoluminescence (PL) map showing the illuminated spot for the two-dimensional PL image profiles b-d) Experimental two-dimensional PL image profiles at b) 0 ns, c) 2.56 ns d) 25.6 ns. The inset axes in d) show the fitted principal diffusion components $D_{x'}$ and $D_{y'}$ of the diffusion tensor for the illuminated grain (where $x'$ and $y'$ are at -34.4 ± 2.2° relative to the images' $x$ and $y$ axes, respectively). Fast carrier diffusion is seen inside the illuminated grain, yet it takes

significantly longer for the second grain to exhibit PL e-g) Simulated two-dimensional PL image profiles corresponding to the same time frames as b-d). White lines denote the identified grain boundaries and are added as a visual aid. b-g) share the same scalebar. h) Time-resolved PL data and optimized simulation result corresponding to the pixel locations of the colored "x"s in b-d, e-g). All images and traces are normalized relative to the maximum pixel value at t = 0 ns.

Given that $MAPbI_3$ has a tetragonal crystal structure at room temperature,[62] anisotropic diffusion is expected and only two parameters are needed to fully describe the principal diffusivities, thus only two independent directions would need to be measured.[22,39] Generally in semiconductors, principal diffusion axes align with the crystallographic axes, though other factors such as restricted diffusion may yield "effective" diffusion coefficients and axes.[22,34,35] Further correlation analyses with structural measurements such as electron-backscatter-diffraction (EBSD) measurements similar to studies by Jariwala *et al.* for PL maps could be used to confirm whether the determined principal axes are true crystallographic axes or effective axes due to restricted diffusion.[39] Indeed, Kelvin probe force microscopy measurements have shown carrier accumulation and potential barriers at grain boundaries which could affect the principal diffusivities.[63] Interestingly, by performing a similar analysis on the adjacent coupled grain, we find that the principal axes in these two separate grains align, with a misorientation angle of 14.3 ± 5.5° (see Fig. S6) which is small relative to crystallographic misorientation previously measured in similar films.[39] This could be an indication that the electronic coupling between grains is dependent on the alignment of the principal axes and crystallographic axes as all other grain boundaries are effectively reflective.[15,20,21] Further extensions of this work could inform preferential crystal growth orientations for higher diffusivity, grain cross-coupling and improved device performance.[17,18,64]

Furthermore, the effect of passivation strategies and cross-linkers on grain boundary transport could be quantified, which can lead to better lateral device design and performance.[25,27,65,66]

## CONCLUSION

In conclusion, we developed a general PDE framework based on a diffusion tensor and material morphology that can accurately describe energy transport in polycrystalline thin films and determine the principal diffusion axes. We show that the PDE model can reproduce measured carrier diffusion in a $MAPbBr_3$ single crystal, where boundary effects are not present and isotropic diffusion is expected. We then apply the model to a polycrystalline $MAPbI_3$ thin film where boundary effects are present and successfully recreate the observed diffusion profiles in terms of shape and intensity. Through this model, we find that illuminated $MAPbI_3$ grains show diffusion anisotropy as expected in tetragonal crystals, with a principal diffusion coefficient of 0.93 $cm^2s^{-1}$ and 0.72 $cm^2s^{-1}$ along the principal diffusion axes. By determining the principal diffusion axes and correlating them with crystal axes or the microstructure, semiconductor crystals can be grown in preferential orientations that maximize the diffusivity along key transport directions. This would allow to minimize barriers to transport in applications where unimpeded transport is paramount such as lateral transport in interdigitated back-contact solar cells and field-effect transistors[67–69]. Furthermore, given the generality of the framework, it can be adapted and correlated with a wide array of multimodal, nanoscale characterization techniques over a range of micro- or nanostructured materials, where anisotropy and boundary effects are present.[70] Experimental and modeling efforts could be further improved at the nanoscale by including time-resolved data generated through electron microscopy, faster light sources and detectors to image faster and smaller spatial changes in energy carrier profiles[13,53,71,72]. This work will likely enable deeper understanding to how crystal structure, orientation, and composition impact local electronic energy transport and loss.

## METHODS

### Materials

Lead iodide ($PbI_2$) was purchased from TCI America (99.99% trace metals basis), Methylammonium iodide (MAI) and methylammonium bromide (MABr) were purchased from GreatCellSolar. Methylammonium chloride (MACl) was purchased from Dyenamo (DN-P08). Methylamine (MA) solution in THF (THF/MA, 2.0M), anhydrous acetonitrile (ACN, 99.8%) and lead (II) bromide ($PbBr_2$) were purchased from Sigma-Alridch.

### Sample Preparation

$MAPbI_3$ thin-films were prepared through a previously reported method using an acetonitrile and THF (THF/ACN) precursor ink solution.[73] 1.15g of PbI2 was weighed out with 0.397g of MAI and 0.0253g MACl under nitrogen in a septum sealed 7mL vial. 2.5mL of THF/MA solution was added through the septum using a Schlenck line to relieve pressure. The solution was stirred at 700 RPM with a 13mm disposable Teflon stir bar until no solids remained. Then 2.5mL of acetonitrile was added to the solution under stirring. This creates a 0.5M (per lead basis) solution with a 1:1:0.15 molar ratio with MAI and MACl. At this point the solution became fully yellow-transparent. The ink was used within one week.

The solution was spincoated on 500μm thick D263 Schott Glass at 2000rpm with a 2000rpm/s ramp for 60 seconds in a dry air (<1% relative humidity) enclosure. The films were then annealed at 100°C for 30 minutes. The samples were subsequently encapsulated in the dry air enclosure with a 0.2mm thick glass coverslip with the use of epoxy (Devcon 20845) and were left to cure at room temperature for 24 hours. This ensured long-term photoluminescence stability during measurements (see Fig. S6).

Growth of $MAPbBr_3$ single crystal thin films: The films were grown using the space-limited inverse-temperature crystallization (ITC) method. Briefly, 1.5 M of $MAPbBr_3$ in DMF was prepared by dissolving an equimolar amount of MABr and $PbBr_2$ in DMF by stirring overnight. ~5 µl of the solution was placed on a glass substrate preheated to the solution temperature and enclosed by another glass substrate. The temperature was then gradually raised from 40°C to 70°C at a rate of 2 °C/hr to induce nucleation and growth. Next, the substrates were separated using a blade and were allowed to cool down to room temperature. Finally, the films were collected for further measurements.

**Time-Resolved Photoluminescence Image Measurements**

Time-resolved microscopy measurements were performed using a modified confocal microscope setup built around a Nikon Eclipse-Ti inverted microscope fitted with an infinity corrected 100× dry objective (Nikon Plan Epi, NA = 0.85) fitted with a correction collar. A fiber-coupled 405nm laser (PicoQuant GmbH, LDH-P-C-405) with pulse duration of <120 ps was collimated with a diameter of ~7.5mm (Thorlabs C40FC-A) and directed towards the objective to excite the sample through the coverslip side. A laser fluence of $15 \mu J/cm^2$/pulse (27.25nW at 500kHz repetition rate with a measured laser spot diameter of 628nm in the x-direction and 652nm in the y-direction) was used unless otherwise stated. The close to diffraction-limited spot ($1/e^2$ diameter of 628nm in the x-direction and 652nm in the y-direction) was focused onto the middle of the perovskite grain being measured to minimize surface light scattering effects. This technique is considered to be far-field, although diffusion properties below the diffraction limit can still be resolved.[32]

The sample emission was filtered through a 405 nm laser bandpass filter set (Chroma ZT405rdc TIRF C157036) along with a 450nm longpass filter (Thorlabs FELH0450). The image was further magnified by 5X through a two-lens telescope (Thorlabs ACH254-040-B followed by a Thorlabs ACH254-200-B) for a total magnification of 500X before being directed to a Micro Photon Devices PDM series single-photon avalanche photodiode with a 50 μm active area on a motorized XY stage with 25mm of travel (Two Thorlabs Z825B Motorized Actuators with a Thorlabs PT3 1" Stage). The resulting 500X image was raster scanned in two dimensions for all the acquired data. The system was calibrated by imaging a 2000 mesh TEM grid (Ted Pella Inc, G2000HS). Time-correlated single photon counting was performed with a PicoHarp 300 and time-resolved photoluminescence images were collected using custom-built acquisition software written in Python based on the ScopeFoundry platform[74].

Intensity photoluminescence images were acquired scanning the laser on the sample through the use of a piezoelectric sample stage (Physik Instrumente, P-733.3CL) which was controlled using a piezo controller (Physik Instrumente, E-710.4CL) with a pixel size of 100 nm and dwell time (integration time) of 10 ms.

**ACKNOWLEGEMENTS**

R.B., D.W.D., and V.B. acknowledge support for this project through the MIT-Tata GridEdge Solar Research Program, which is funded by the Tata Trusts. R.B. acknowledges support from the National Science Foundation Graduate Research Fellowship under Grant No. (1122374). R.B. acknowledges support from MathWorks through the MathWorks Engineering Fellowship. A.Y.A. and O.M.B acknowledge the financial support provided by King Abdullah University of Science and Technology (KAUST). R.B. thanks Samuel D. Stranks (University of Cambridge), William A. Tisdale (MIT), Mikhail M. Glazov (Ioffe Insitute) and Sarthak Jariwala (Xerox PARC) for helpful discussions.

Supporting Information for

# Mapping the Diffusion Tensor in Microstructured Perovskites

*Roberto Brenes,[1,2] Dane W. deQuilettes,[1*] Richard Swartwout,[1] Abdullah Y. Alsalloum,[3] Osman M.Bakr,[3] Vladimir Bulović[1,2*]*

[1] Research Laboratory of Electronics, Massachusetts Institute of Technology, 77 Massachusetts Avenue, Cambridge, Massachusetts 02139, USA

[2] Department of Electrical Engineering and Computer Science, Massachusetts Institute of Technology, 77 Massachusetts Avenue, Cambridge, Massachusetts 02139, USA

[3] Division of Physical Sciences and Engineering, KAUST Catalysis Center (KCC), King Abdullah University of Science and Technology, Thuwal 23955-6900, Kingdom of Saudi Arabia

*Corresponding Authors: danedeq@mit.edu; bulovic@mit.edu

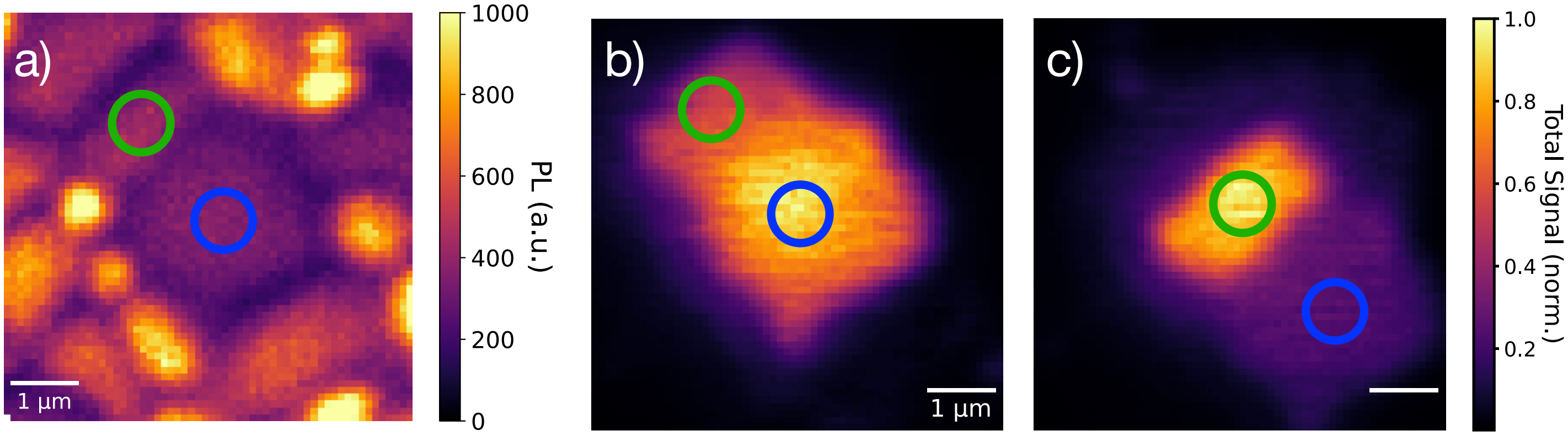

**Figure S1.** a) Photoluminescence (PL) map of the $MAPbI_3$ shown in Figure 3 of the main text, with colored circles indicating the large (blue circle) and small (green circle) grains illuminated in the widefield PL images obtained in b) and c). b) Widefield PL image where the large (blue) grain is illuminated, showing electronic coupling to the small (green) grain. c) Widefield PL image where the small (green) grain is illuminated, showing electronic coupling to the large (blue) grain. b) and c) share the same scalebar.

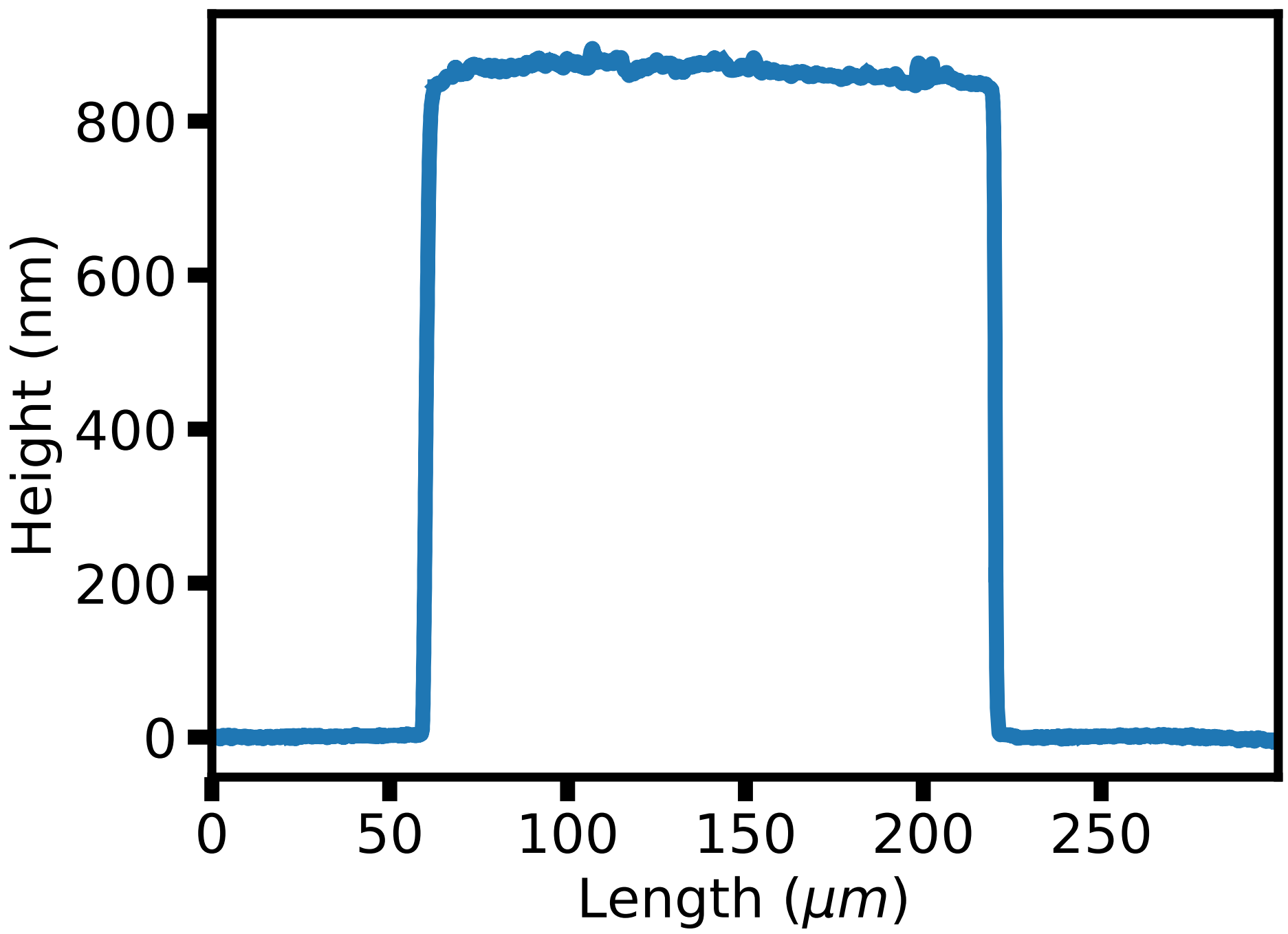

**Figure S2.** Profilometry scan of a portion of the $MAPbBr_3$ single crystal. The thickness of the crystal is roughly 880nm as determined from the step between the substrate and the sample.

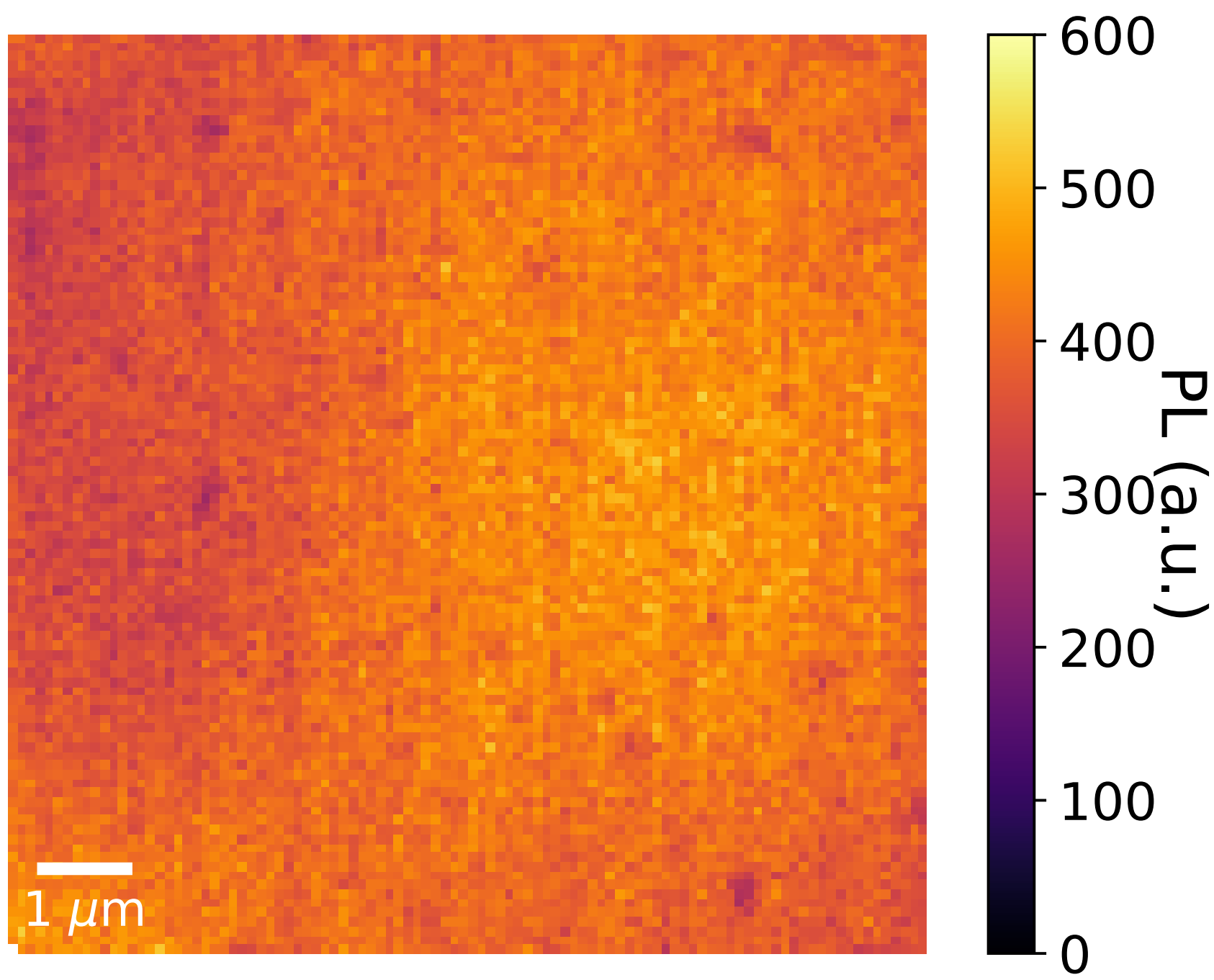

**Figure S3.** 10 μm by 10 μm photoluminescence map of the single crystal $MAPbBr_3$ thin-film shown in the main text. Little heterogeneity is observed, indicating a high quality crystal with no discernible interfaces.

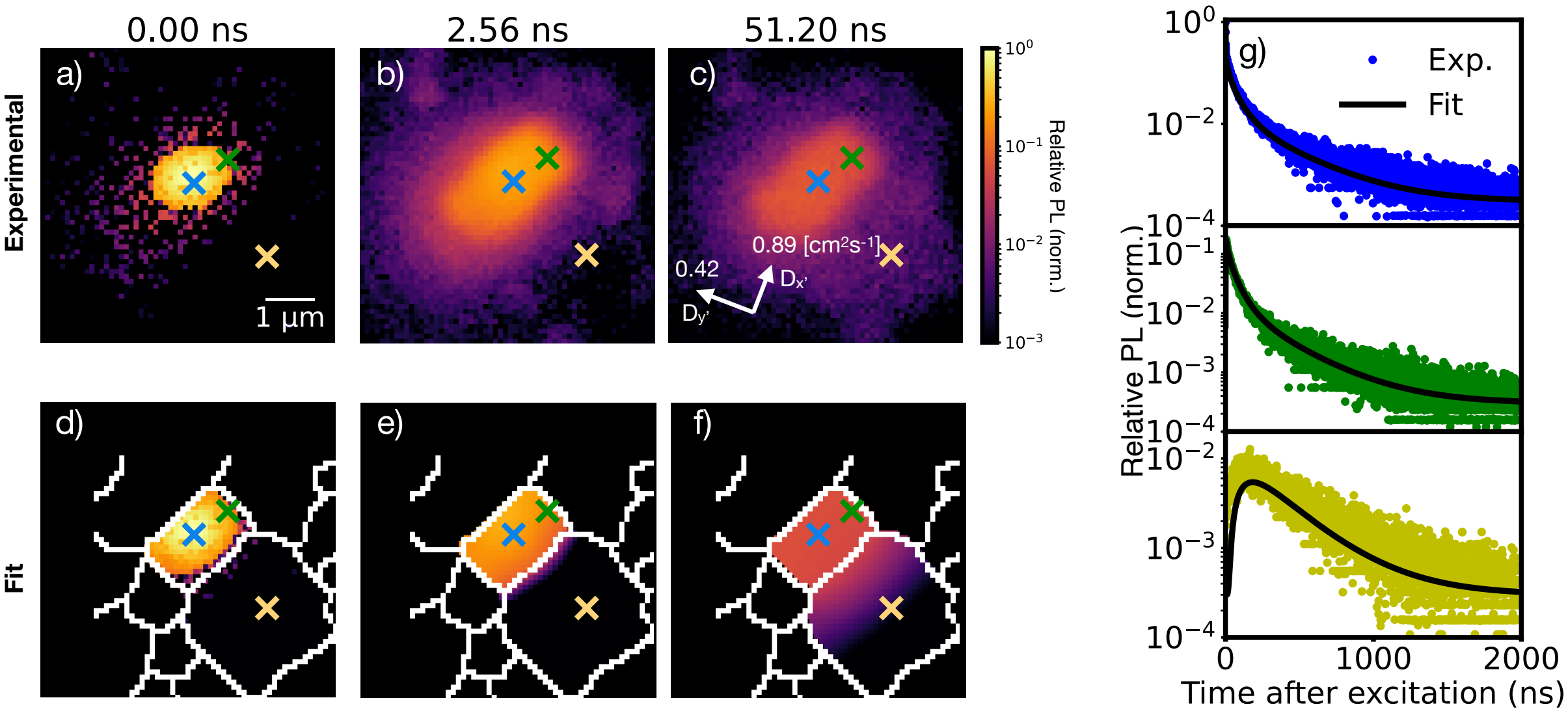

**Figure S4.** Two-dimensional photoluminescence images when illuminating the smaller, electronically coupled grain in Figure 3 of the main text**.** a-c) Experimental two-dimensional photoluminescence (PL) image profiles at a) 0 ns, b) 2.56 ns c) 51.20 ns. The inset axes in c) show the fitted principal diffusion components $D_{x'}$ and $D_{y'}$ of the diffusion tensor for the illuminated grain (where *x'* and *y'* are at 69.1° relative to the images' *x* and *y* axes, respectively). Fast carrier diffusion is seen inside the illuminated grain, yet it takes significantly longer for the second grain to exhibit PL, similar to Figure 3 in the main text. d-f) Simulated two-dimensional PL image profiles corresponding to the same time frames as a-c). White lines denote the identified grain boundaries and are added as a visual aid. a-f) share the same scalebar. g) Time-resolved PL data and optimized simulation result corresponding to the pixel locations of the colored crosses in a-c, d-f). All images and traces are normalized relative to the maximum pixel value at t = 0 ns.

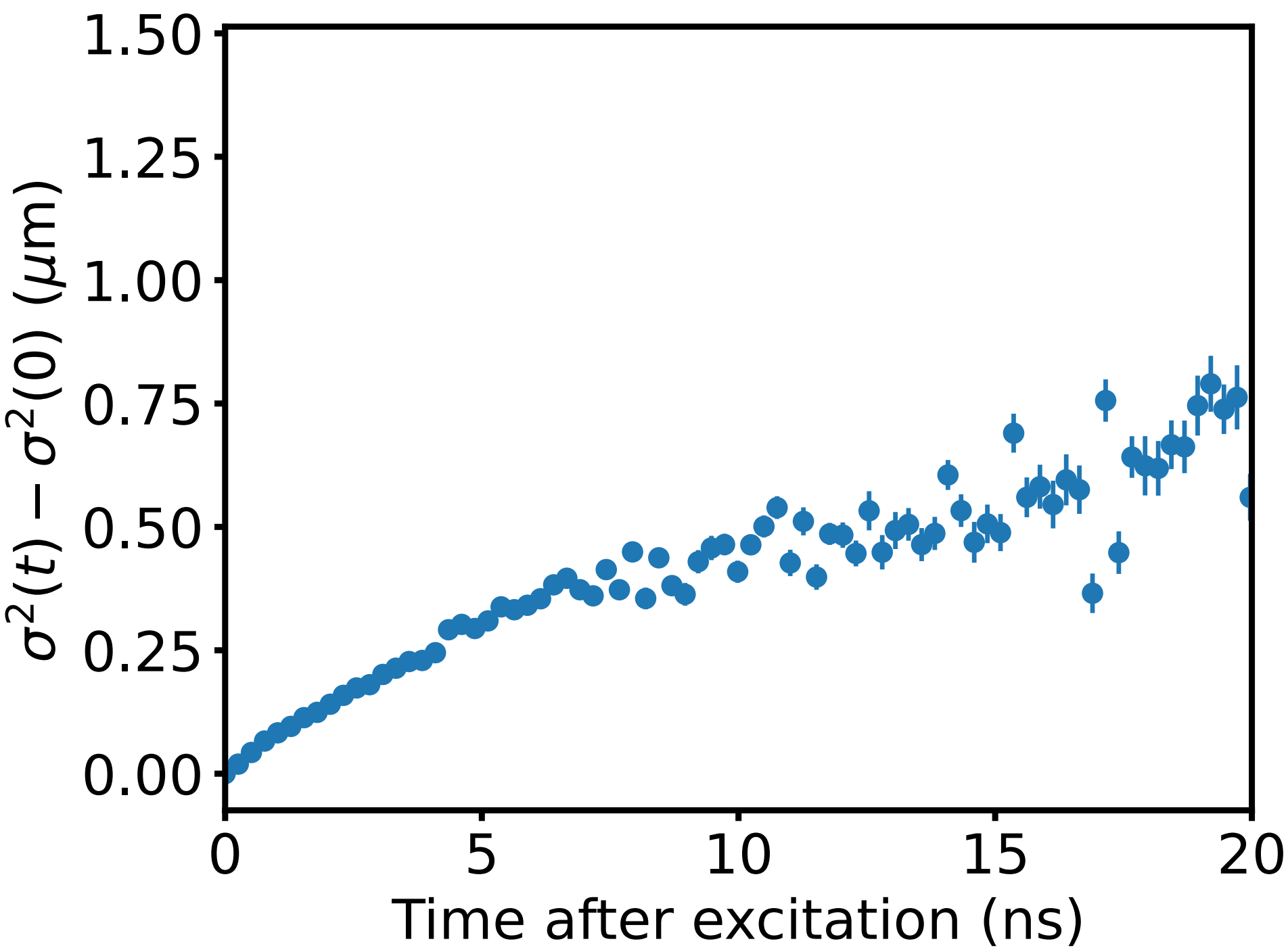

**Figure S5.** Mean-squared-displacement ($\sigma^2(t) - \sigma^2(0)$) profile in the x-direction for the carrier diffusion data presented on Figure 2 of the main text. $\sigma^2$ is the Gaussian variance of the fitted Gaussian profile to the normalized PL intensity profile. Since the sample is homogeneous, a change in the slope of the variance may indicate a time-dependent diffusion coefficient as explained in our previous work.[1]

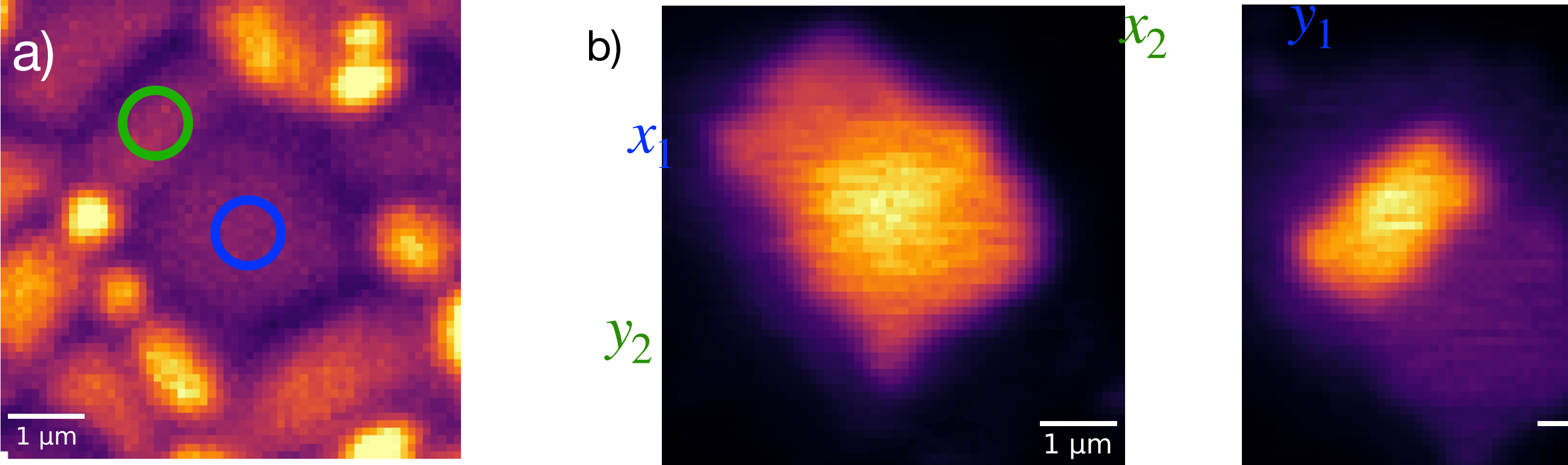

**Figure S6.** a) Photoluminescence (PL) map showing the large (blue) and small (green) grains illuminated for the diffusion measurements in Fig. 3 of the main text and Fig. S4. b) Diagram showing the angular misorientation between the principal diffusion axes of the large grain (blue) and small grain (green). The principal diffusion axes of the are at an angle of -34.4° and 69.1° relative to the image *x* and *y* axes for the large grain and small grain, respectively.

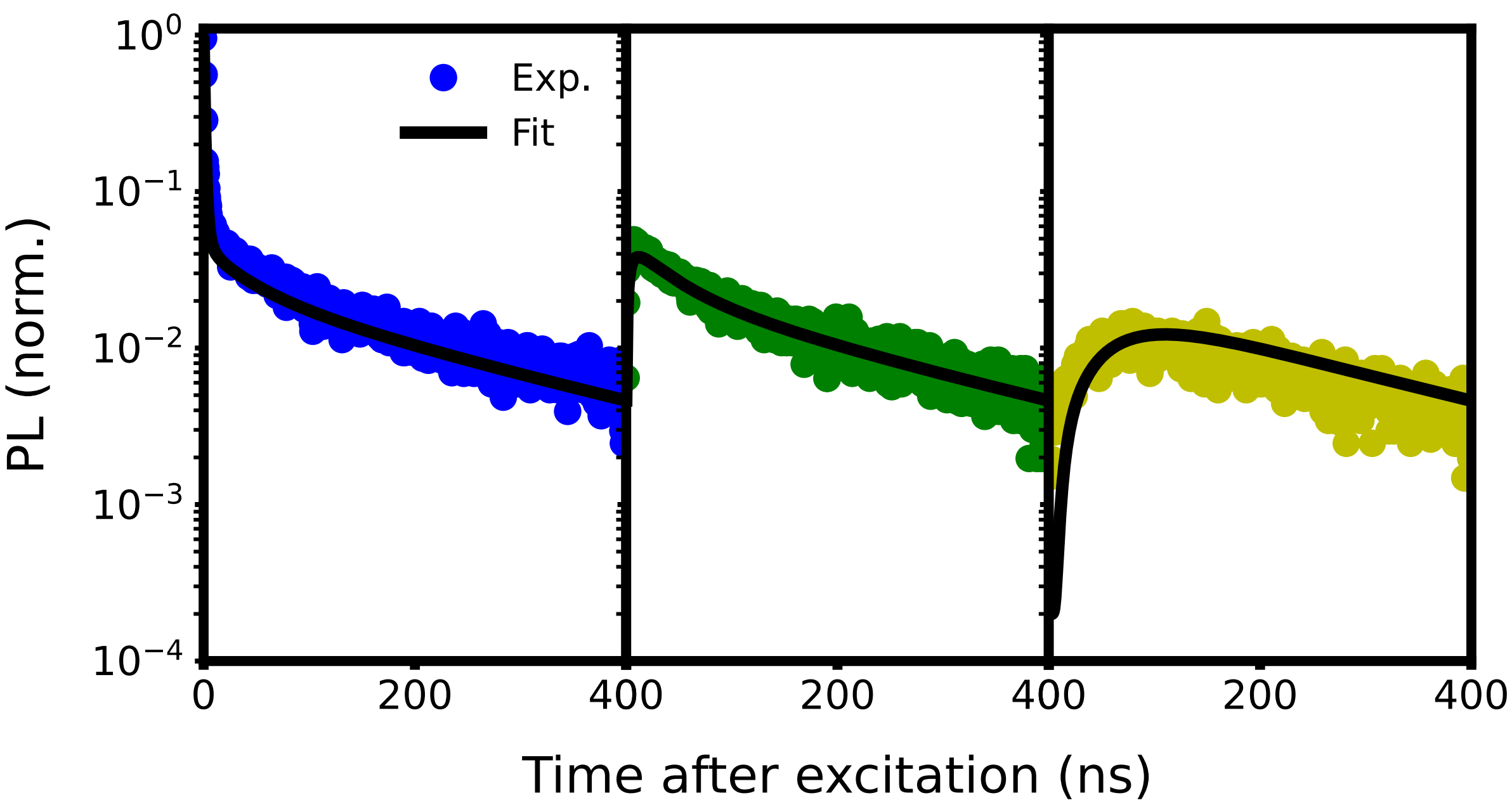

**Figure S7.** Time-resolved PL data and optimized simulation result corresponding to the same traces shown in Fig. 3h) of the main text shown on an expanded timescale (i.e. shorter time window).

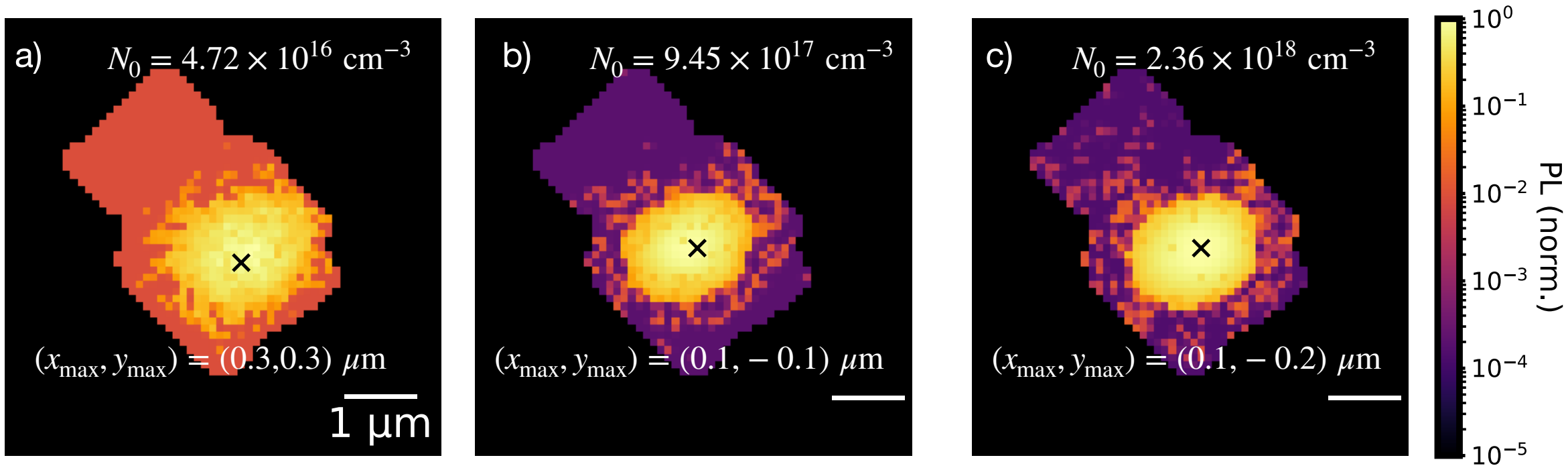

**Figure S8.** Initial photoluminescence frame for initial carrier densities a) $N_0 = 4.72 \times 10^{16}$cm$^{-3}$, b) $N_0 = 9.45 \times 10^{17}$cm$^{-3}$. c) $N_0 = 2.36 \times 10^{18}$cm$^{-3}$ where luminescence outside of the geometry has been thresholded to zero. The corresponding excitation center coordinates $(x_{max}, y_{max})$were determined by where the maximum photoluminescence occurs at t=0. The “x” marks the location of the excitation center.

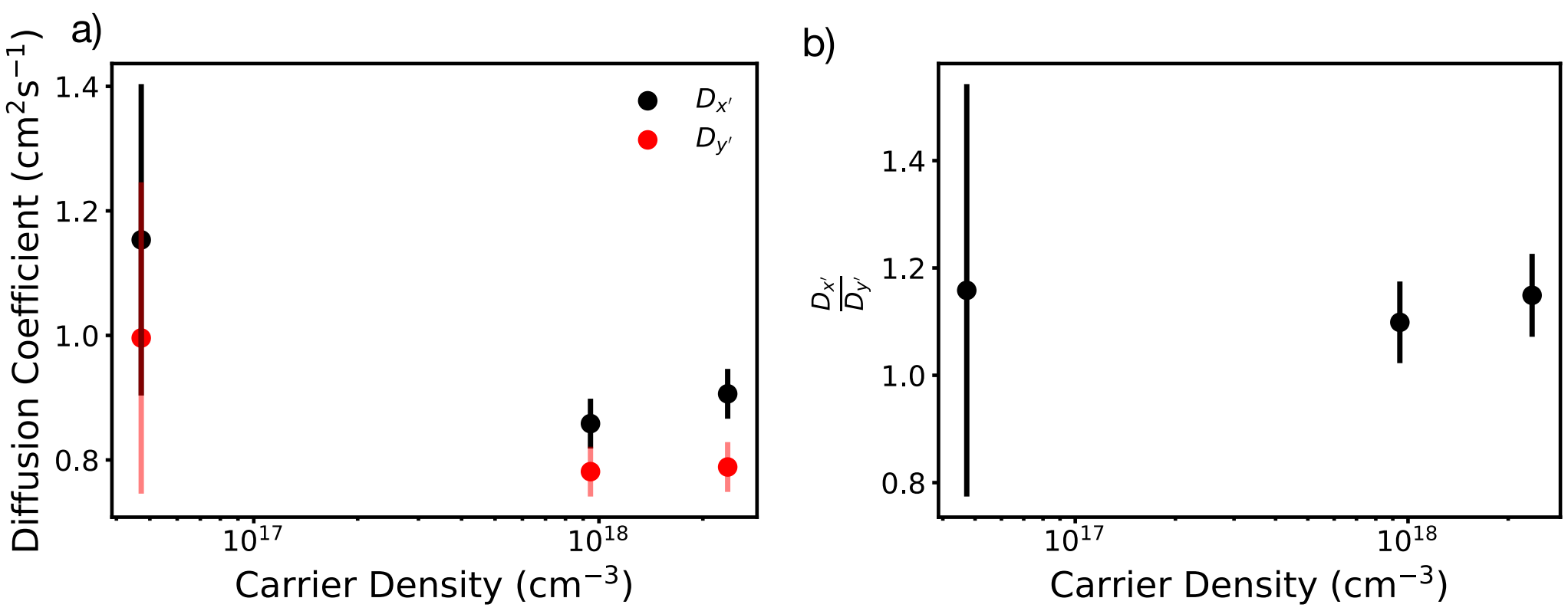

**Figure S9.** a) Fitted principal diffusion coefficients $D_{x\prime}$ and $D_{y\prime}$ along with their 95% confidence interval as a function of carrier density for the same illuminated grain as in Fig. 3 of the main text. Wider confidence intervals for carrier densities <$10^{17}$ $cm^{-3}$ are due to low signal to noise ratio b) Ratio of $D_{x\prime}/D_{y\prime}$ as a function of carrier density for the same illuminated grain as in Fig. 3 of the main text, which shows consistent agreement of $D_{x'} > D_{y'}$ over the carrier densities probed.

## S1. Modeling Spatially and Time-Resolved Carrier Diffusion

As discussed in the main text, the phenomenological model used for modeling carrier diffusion with anisotropy is as follows:

$$\frac{\partial \mathrm{N}(\boldsymbol{u},t)}{\partial t} = \nabla \cdot (\boldsymbol{D}(\boldsymbol{u})\nabla \mathrm{N}(\boldsymbol{u},t)) - k_1 \mathrm{N}(\boldsymbol{u},\mathrm{t}) - P_{esc} k_2 \mathrm{N}^2(\boldsymbol{u},t) - k_3 \mathrm{N}^3(\boldsymbol{u},t) \qquad \text{(S1)}$$

where $\boldsymbol{D}$ is a 2x2 diffusion tensor; $\nabla$ is the gradient operator; $\mathrm{N}(\boldsymbol{u},t)$ is the carrier density as a function of position $\boldsymbol{u} = (x,y)$ in the measurement axes and time; $k_1$ is the non-radiative, first-order (monomolecular) recombination constant, $k_2$ is the second-order (bimolecular) recombination rate constant, $k_3$ is the third-order (Auger) recombination rate constant and $P_{esc}$ the average probability of escape for a photon emitted inside the film. Kiligaridis *et al.* have observed that the ABC recombination model used in our simulations does not capture the recombination dynamics of lead halide perovskites in the quasi-CW regime at low (<0.1 suns) excitation regime but works well for uncoated perovskite films in the high carrier density (> 1 suns) single pulse regime[2]. Our experiments were conducted in the pulsed regime with high initial photoexcited carrier concentrations >$10^{17}$ $cm^{-3}$ (>100 suns) and photoluminescence noise floors with carrier densities ~ $10^{16}$ $cm^{-3}$, where the ABC model can be applied and where trap-related processes are likely saturated[2–5]. Furthermore, we have ignored subnanosecond processes such as hot carrier transport and rapid carrier trapping which can occur on the timescale of ~100 ps, since our data has a minimum time resolution of 0.511 ns above which recombination and diffusion processes are more dominant[6,7]. Ion migration effects were not included in this model since these processes occur on the timescale of multiple seconds, which is significantly longer than the measurement timescale[8–10].

As mentioned in the main text, the diffusion tensor is typically described by a second-rank tensor of the form:

$$\boldsymbol{D} = \begin{pmatrix} D_{xx} & D_{xy} \\ D_{yx} & D_{yy} \end{pmatrix} \quad \text{(S2)}$$

where the diagonal terms, $D_{xx}$ and $D_{yy}$, are the diffusivities in the $x$ and $y$ directions in the measurement coordinates, respectively; and off-diagonal terms, $D_{xy}$ is the diffusivity along the $x$ direction arising from a concentration in the $y$ direction and $D_{yx}$ is the diffusivity along the $y$ direction arising from a concentration in the $x$ direction. Since the tensor is a second-rank tensor, it must have conjugate symmetry, that is, the off-diagonal elements must be equal to each other $D_{xy} = D_{yx}$ [11,12].

Any second-rank tensor can be converted to its principal axes and thus principal diffusivities. These axes typically correspond to the crystallographic axes, though other effects such as restricted diffusion due to boundaries or inhomogeneities may influence their orientation and the principal diffusivities can turn into effective principal diffusivities that do not align with the crystallographic axes[11,12]. The measurement axes $x$ and $y$ must be rotated by some angle θ to the principal axes $x'$ and $y'$, which can be achieved through the rotational matrix R:

$$\mathrm{R} = \begin{bmatrix} \cos\theta & -\sin\theta \\ \sin\theta & \cos\theta \end{bmatrix} \quad \text{(S3)}$$

In order to determine the appropriate rotation angle θ, we can diagonalize the diffusion tensor $\boldsymbol{D}$ by determining it's eigenvalues $\lambda_1$, $\lambda_2$ and eigenvectors $\widehat{e_1}$, $\widehat{e_2}$. The diagonalized second-rank diffusion tensor $\boldsymbol{D'}$ will be:

$$\boldsymbol{D'} = \begin{bmatrix} \lambda_1 & 0 \\ 0 & \lambda_2 \end{bmatrix} = \begin{bmatrix} D_{x'} & 0 \\ 0 & D_{y'} \end{bmatrix} = R \cdot \boldsymbol{D} \cdot R^T \tag{S4}$$

where the eigenvalues correspond to the principal diffusivities $D_{x'} = \lambda_1$ and $D_{y'} = \lambda_2$ , $R^T$ is the transpose of $R$, and the columns of $R$ consist of the eigenvectors $\widehat{e_1}$ and $\widehat{e_2}$:

$$\mathrm{R} = \begin{bmatrix} e_x^1 & e_x^2 \\ e_y^1 & e_y^2 \end{bmatrix} \tag{S5}$$

$e_x^i$ and $e_y^i$ are the $i$-th eigenvector's $x$ and $y$ component respectively. To finally find the angle θ between the measurement and principal axes, we can equate Eqs. S3 and S5 to get:

$$\theta = \cos^{-1}(e_x^1) = \sin^{-1}(e_y^1) = -\sin^{-1}(e_x^2) = \cos^{-1}(e_y^2) \tag{S6}$$

The average photon probability of escape $P_{esc}$ used for all the simulations was 14%. This value is based on the calculation by Pazos-Outón *et al.* and our previous work, where a probability of escape of 7% was calculated for a film with a perfect back-reflector[13,14]. Since our film does not contain a back-reflector, we have doubled the probability of escape since photons can now escape through either the front or back surface of the film. The fitted monomolecular recombination rate constant $k_1$ was kept the same throughout the entire simulation geometry, even across different grains (see Section S4). The bimolecular recombination rate constant was fixed at $k_2 = 2 \times 10^{-10}$ cm$^3$s$^{-1}$ and the Auger recombination rate constant was fixed at $k_3 = 1 \times 10^{-29}$ cm$^6$s$^{-1}$ as has been widely reported in literature[15–19]. We note that $k_2$ is an intrinsic material parameter which appears to vary in literature due to the wide-ranging values of $P_{esc}$ which is why we have explicitly incorporated $P_{esc}$ into our model to keep $k_2$ consistent with literature reported values which correct for photon reabsorption [15,16]. Furthermore, we have also fixed $k_3$ since its value depends on the intrinsic energy band properties of the material, our fitted error is independent of its value (see Fig. S15) and reported literature values lie within a small range ($1 \times 10^{-28} > k_3 > 1 \times 10^{-30}$ $cm^6s^{-1}$)[20,21]

The partial differential equation (PDE) was implemented in two dimensions in MATLAB using the PDE Toolbox. The non-linear least squares minimization was carried out in MATLAB as well through the use of the optimization toolbox. To ensure that the minimization scheme led to a global minimum, the Global Optimization Toolbox was used to solve multiple minimization problems with 50 different starting points for the initial parameter guesses. The multiple optimization problems were run in parallel on an Amazon EC2 AWS server (m5.12xlarge, 24 core) with a headnode machine (c5d.xlarge, 2 core, 1x100 NVME).

## S2. Initial Condition for PDE Model

The initial condition for the carrier distribution $N(x, y, t = 0)$ was determined through the use of the initial frame from the time-resolved PL image. Since PL ~ $N^2$ then we can determine the initial carrier distribution by:

$$N(x, y, t = 0) = A\sqrt{PL(x, y, t = 0)} \tag{S7}$$

where $A$ is a proportionality factor between the carrier density and the square root of the PL profile given by:

$$A = \frac{N_0}{d \int_x \int_y \sqrt{PL(x,y,t=0)}dxdy} \tag{S8}$$

where we have distributed the total number of carriers excited in the film $N_0$ throughout the square root of the PL intensity profile and the thickness of the film $d$. We can determine the total number of carriers excited in the film by considering the amount of laser photons being absorbed by the film, which is given by:

$$N_0 = \frac{P_\lambda}{RepRate} \times \frac{\lambda}{hc}(1 - 10^{-OD}) \quad \text{(S9)}$$

where $P_\lambda$ is the incident laser power, λ is the laser wavelength, h is Planck's constant, c is the speed of light, $RepRate$ is the laser pulse repetition rate and $OD$ is the film's optical density. This initial condition method is similar to others reported in literature.[22,23]

**S3. Microstructure Recognition and Boundary Conditions**

In order to create the PDE geometry for the simulations, we used an edge-finding algorithm from MorphoLibJ on measured PL maps of the area surrounding the illuminated grain in the time-resolved PL images[24]. Figure S1 shows the determined edges from the PL map of the same area captured in the time-resolved PL images shown in Figure 3 in the main text. Only the grains that were determined to be electronically coupled were finally used in the simulation (see Fig. S1).

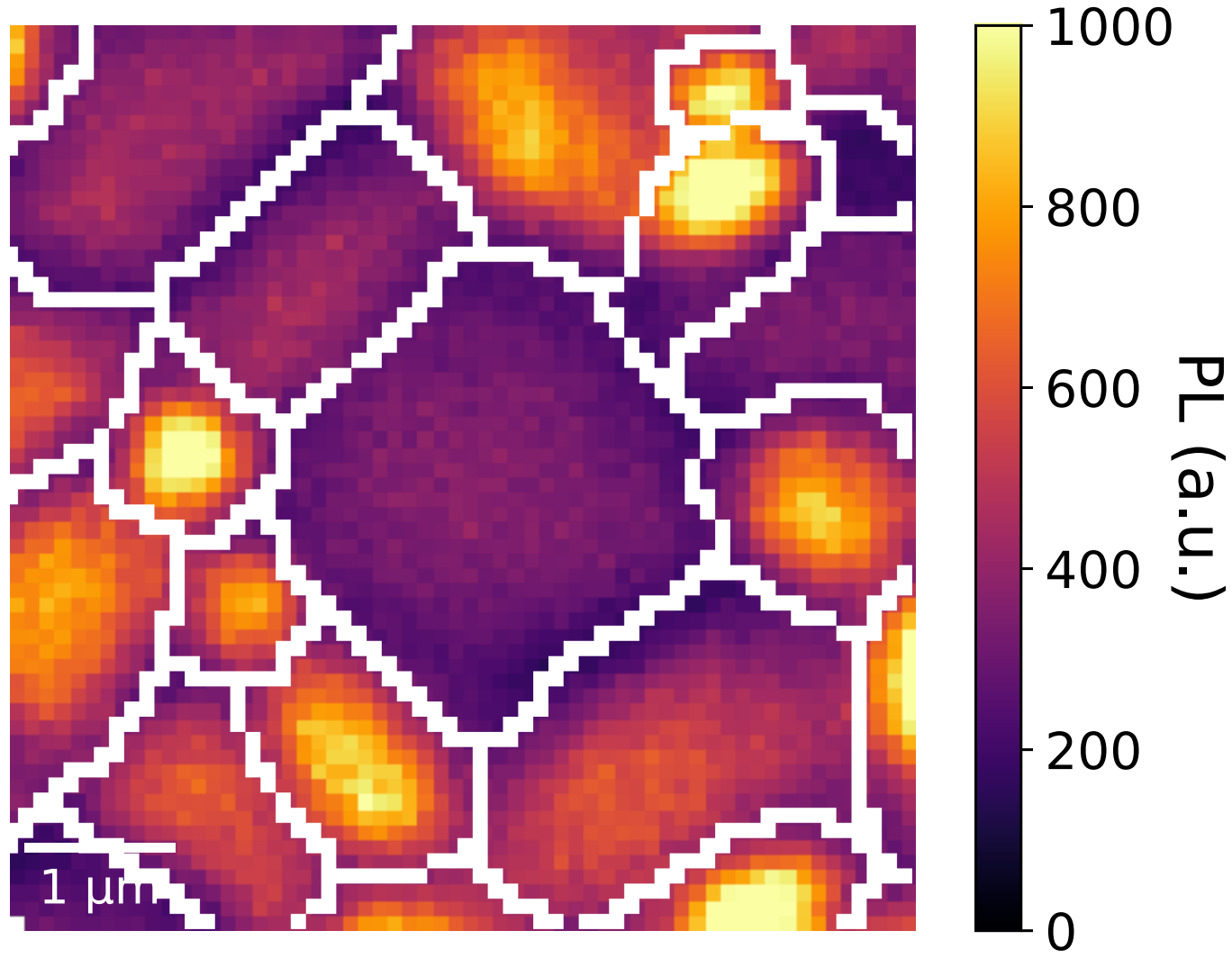

**Figure S10.** 5 µm by 5 µm photoluminescence (PL) map with the identified grain edges overlaid in white.

To turn these edges into a closed polygon, the boundaries were binarized, skeletonized and traced. The resulting boundaries were dilated and then traced for each inner boundary. The resulting x and y coordinates of each traced line segment was then used to define closed polygons for the PDE model.

The domain boundaries that were not adjacent to each other were considered to have no flux (i.e. reflective) boundary conditions, as has been shown to be the case in lead halide perovskites in previous work.[22,25–28] This was implemented following generalized Neumann boundary conditions of the form:

$$\left.\frac{\partial N(u,t)}{\partial x}\right|_{\gamma} = \left.\frac{\partial N(u,t)}{\partial y}\right|_{\gamma} = 0 \qquad \text{(S10)}$$

where **γ** is the vector that denotes the spatial coordinates of the grain boundaries. Since Matlab's PDE Toolbox utilizes a finite-element method based on the weak formulation of PDEs, no boundary conditions can be specified between subdomain boundaries, even if the coefficients are discontinuous[29].

## S4. Validity of a Single $k_1$ Value for a Single Film

As mentioned in Section S1, the monomolecular recombination rate constant $k_1$ was kept constant throughout the entire film, even across different grains. By looking at PL maps, where the laser is scanned on the sample, it is possible to see intensity fluctuations between grains (see Fig. S1), which one could attribute to local changes in the $k_1$ value. However, we posit that these changes in intensity are due to charge carrier diffusion effects, rather than changes in the monomolecular recombination rate constant. Indeed, as we have shown before, there is a negative correlation between the intensity of a grain and its size (i.e. the larger the grain, the lower the PL intensity).[25] To investigate this further, we performed PL maps on the $MAPbI_3$ with a focused illumination spot. We then enlarged the beam to a diameter of ~10 $\mu$m by focusing the laser spot to the back aperture of the microscope objective. By creating such a large spot, the excited carrier density, especially near the center of the illumination spot, should lead to a relatively flat carrier density, minimizing the amount of diffusion. We can then selectively collect PL from the middle of the excitation spot by centering the detector on the illumination spot. Since the detector has an active area of 50 μm and the image is being magnified by 500X, the detector acts as a pinhole with an effective diameter of 100nm in the sample scale.

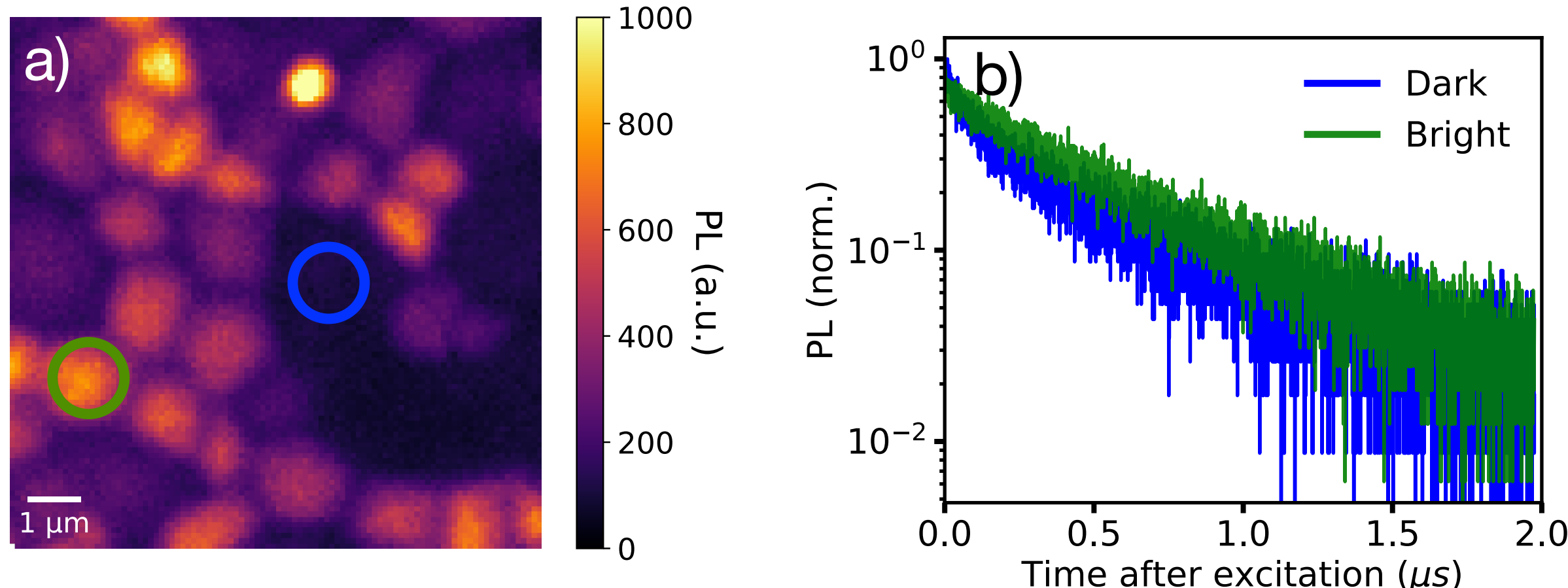

**Figure S11.** a) Photoluminescence (PL) map with circled locations where the detector was centered to obtain time-resolved PL (TRPL) traces b) Dark (blue) and bright (green) grain TRPL traces of the corresponding colors of the areas shown in a) measured with a laser spot of ~10µm to minimize diffusion effects on the traces. Both traces show an almost identical long decay component for both the dark and the bright grain in the PL map.

Figure S2a shows the PL map obtained for the $MAPbI_3$ with the focused laser illumination, where one can see typical PL heterogeneity between grains. Two different grains were identified, a "bright" grain (green circle) and a "dark" grain (blue circle) where time-resolved PL (TRPL) traces would be measured with the enlarged illumination spot. Figure S2b shows the collected TRPL traces for both grains using the same corresponding colors to highlight them in Figure S2a. The TRPL traces exhibit an almost identical single exponential long decay component, indicating similar recombination dynamics. As stated before, under these measurement conditions, very little to no diffusion effects should be present, therefore most of the dynamics should be due to charge carrier recombination. Since a relatively low fluence was used ($<0.1 \mu J/cm^2$/pulse), the TRPL dynamics should be mostly dominated by monomolecular recombination. Given their similar shape, we conclude that the difference in the monomolecular recombination rate constant between the grains must be minimal.

**S5. Removing Waveguided and Scattered PL**

As reported by others in literature[30–32], we found that the measured PL images showed a component that we attribute, as others did, to scattered or waveguided light. This is because the absorption length of re-emitted photons in lead-halide perovskites is typically ~1 μm[14,30], while the scattered component can be detected at distances far greater than that. Figure S3 shows a TRPL trace collected 15 μm away from the excitation spot in the $MAPbBr_3$ thin-film single crystal discussed in the main text. A fast decay at t=0 ns is detected, even when most re-emitted photons would have been absorbed at these lengths. As we discussed in our previous work[1], most re-emitted photons occur at time-scales in the order of the radiative lifetime of the material (~100 ns for $1x10^{17}cm^{-3}$), therefore this effect should mostly be due to scattered light from non-uniformities or roughness in the crystal.

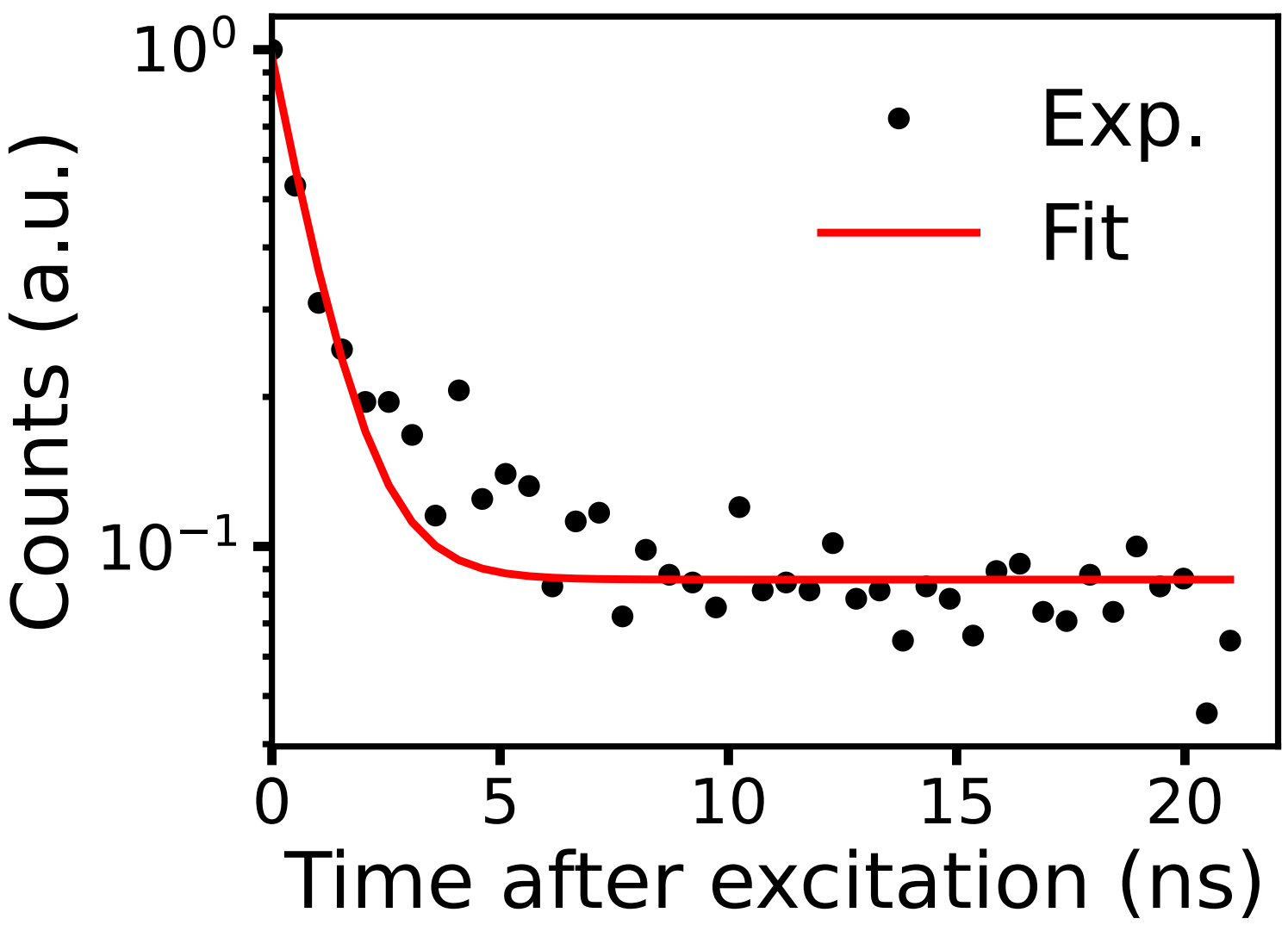

**Figure S12.** Measured TRPL data (black) approximately 15 μm away from the illumination spot. The red line shows the exponential fit to the data, with a fitted lifetime of about 0.87ns.

Since our model does not consider scattered or waveguided photons, we subtract this contribution to our data for both $MAPbBr_3$ and $MAPbI_3$ thin films. In order to do so, we fit an exponential of the form:

$$PL_{WG} = A\,exp(-t/\tau) + B \tag{S11}$$

where $PL_{WG}$ is the waveguided PL signal, $\tau$ is the lifetime of the decay, $t$ is time, $A$ is intensity and $B$ the background counts. In our case, we found $\tau = 0.87$ ns, $A = 0.88$ and $B = 0.085$. This exponential was then matched to the intensity at $t = 0$ ns for the pixels where the waveguided component was removed.

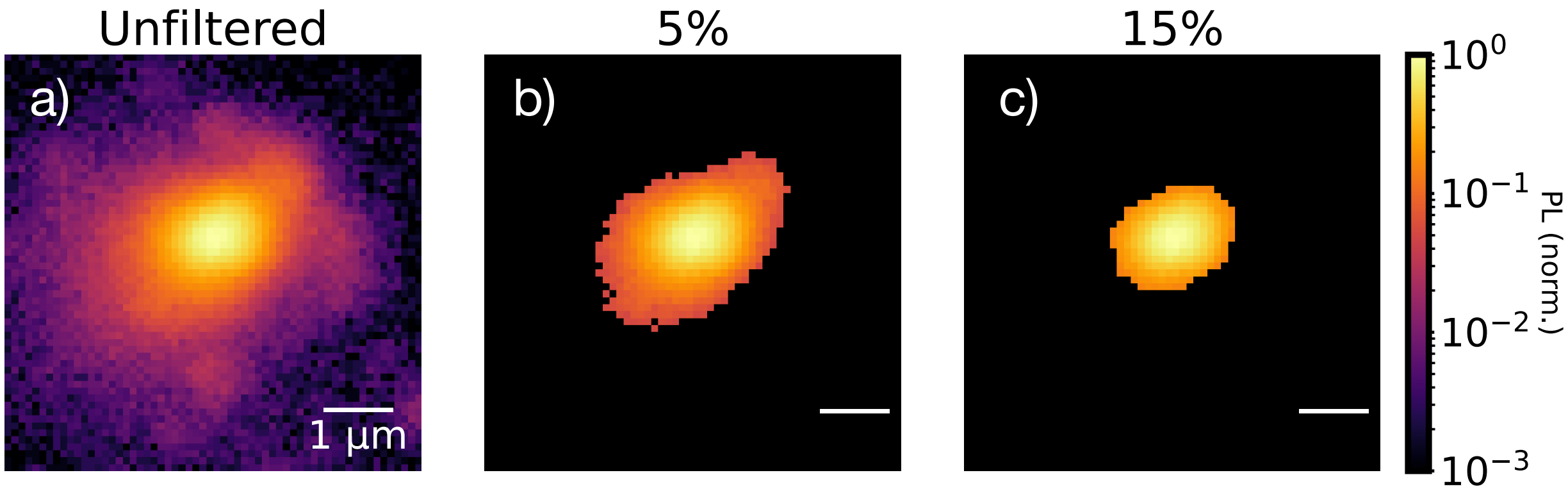

**Figure S13.** Initial frame of photoluminescence image of a $MAPbI_3$ thin-film shown in the main text in Fig. 3 with a) no removal of waveguided/scattered component b) removal of waveguided/scattered component where the intensity falls to 5% of the maximum image intensity c) removal of waveguided/scattered component where the intensity falls to 15% of the maximum image intensity. All images share the same scalebar.

To determine which pixels would have the waveguided component removed, we explored different threshold values of the maximum intensity of the time-resolved PL images. Figure S4a shows the raw data at $t$ = 0 ns for the time-resolved data presented in Figure 3 in the main text. Even though the film was excited with a focused 405nm laser spot with a spot size ~620nm, we see significant PL signal away from the excitation spot due to scattered PL. Figures S4b and S4c show the same image but where the scattered component was subtracted at pixels where the intensity fell to 5% and 15% of the maximum PL value, respectively. Since we expect most of the PL at $t$ = 0 ns to be due to recombining excited carriers from the laser excitation spot, this profile should be as close as possible to the shape of the laser spot, therefore the 15% threshold was chosen.

## S6. Fitting Error Estimation

To test the accuracy of our fitting model, we decided to fit a simulated PL profile according to Eq. S1. In this case, the $k_1$ monomolecular recombination rate constant was fixed at $1 \times 10^6$ $s^{-1}$ while the diffusion tensor $\boldsymbol{D_{sim}}$ was set to:

$$\boldsymbol{D_{sim}} = \begin{pmatrix} D_{xx} & D_{xy} \\ -D_{yx} & D_{yy} \end{pmatrix} = \begin{pmatrix} 1.5 & 0.3 \\ 0.3 & 0.5 \end{pmatrix} cm^2 s^{-1} \tag{S12}$$

We used an airy disk excitation profile as the initial condition, since the excitation profile for a diffraction-limit spot in confocal microscopy is often described as an airy disk. The functional form of an elliptical Airy disk excitation profile is given by:

$$G(r, 0) = \frac{N_0}{I_0} \left( \frac{2J_1(1.22\pi r)}{1.22\pi r} \right)^2 \tag{S13}$$

Where $J_1$ is the Bessel function of the first kind, $r = \sqrt{\left(\frac{x}{x_0}\right)^2 + \left(\frac{y}{y_0}\right)^2}$, $x_0$ is the *x*-coordinate along the *x*-axis where the first zero occurs, and $y_0$ is the y-coordinate along the *y*-axis where the first zero occurs, $I_0 = \iint_A B(r)dxdy$ where $B(r) = \left( \frac{2J_1(1.22\pi r)}{1.22\pi r} \right)^2$ and A is the domain area of the simulation. We set $N_0 = 1 \times 10^{17}\ cm^{-3}$ and $x_0 = y_0 = 300\ nm$.

The simulated PL profiles are shown in Figure S5. Figures S5a,b show the initial frame at t = 0 ns and a frame at t = 5.12 ns. Figure S5b clearly shows significant diagonal smearing due to the non-zero off-diagonal diffusion tensor component. Figure S5c shows selected traces in the locations specified by the colored circles in Figure S5b. Significant diagonal asymmetry is present due to the non-zero off-diagonal diffusion tensor component.

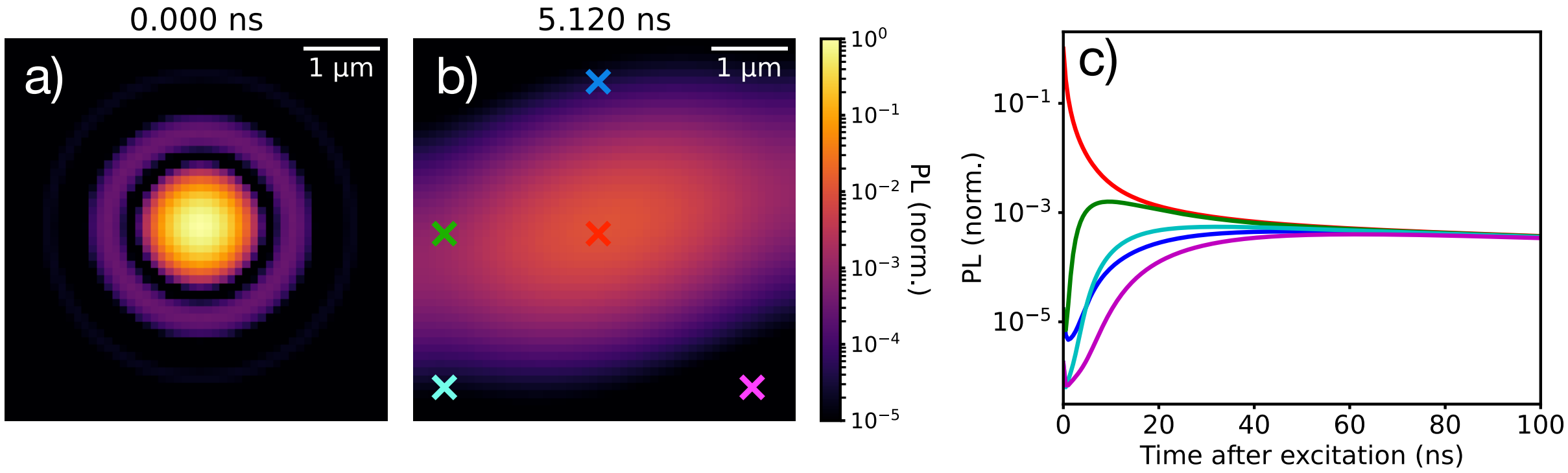

**Figure S14.** Simulated photoluminescence (PL) images using Eq. S1 shown at frames at a) 0 ns and b) 5.12 ns. An airy disk initial condition was used for the simulated laser spot excitation profile. c) Simulated time-resolved PL traces corresponding to the pixel locations of the colored "x"s shown in b).

The simulated data was then fit through a global non-linear least squares minimization as in the main text. The fitted diffusion tensor $\boldsymbol{D}_{fit}$ along with the 95% confidence interval as the uncertainty, was**:**

$$\boldsymbol{D}_{fit} = \begin{pmatrix} 1.5120 \pm 0.0001 & 0.3026 \pm 0.0002 \\ 0.3026 \pm 0.0002 & 0.5118 \pm 0.0001 \end{pmatrix} \mathrm{cm}^2\mathrm{s}^{-1} \tag{S14}$$

The error for the non-zero diffusion tensor components is small, where the fitted $D_{xx}$ has an error of 0.8%, $D_{xy} = D_{yx}$ an error of 0.87% and $D_{yy}$ an error of 2.36%. However, the 95% confidence intervals represent an uncertainty that is too small, given that the original values do not fall between that uncertainty range. We do not know what other uncertainty sources (i.e. numerical error, initial condition discretization, PDE numerical error) might not be captured by the 95% confidence interval determined from the Jacobian. However, we determined that a change of ~1% in the minimized residual due to dependent variations in the diffusion tensor was enough for the uncertainty make the fitted value fall in the range of the input diffusion tensor $\boldsymbol{D_{sim}}$. We then chose a more conservative residual value change of 5% for the minimized residuals due to changes in optimized fitted values as the uncertainties for the fitted experimental values.

**S7. Fitted Coefficients**

Fitted model parameters, along with their error determined using the method above, are reported in the tables below for each dataset. Note that parameters where the uncertainty include zero indicate that the value is not significantly different from zero in the model.

**Table S1: Fitted coefficients $MAPbBr_3$ Single Crystal (Figure 2)**

| Diffusion Tensor [$cm^2s^{-1}$] | $k_1$ [$\times \mathbf{10^6}$ $s^{-1}$] |
|---|---|
| $\begin{bmatrix} \mathbf{0.95 \pm 0.03} & \mathbf{0.0000 \pm 0.001} \\ \mathbf{0.0000 \pm 0.001} & \mathbf{0.93 \pm 0.03} \end{bmatrix}$ | $\mathbf{3.34 \pm 0.01}$ |

**Table S2: Fitted coefficients polycrystalline $MAPbI_3$, illuminated large grain (Figure 3)**

| Diffusion Tensor [$cm^2s^{-1}$] | Diffusivity (coupled grain) [$cm^2s^{-1}$] | $k_1$ [$\times \mathbf{10^6}$ $s^{-1}$] |
|---|---|---|
| $\begin{bmatrix} \mathbf{0.86 \pm 0.02} & \mathbf{0.10 \pm 0.02} \\ \mathbf{0.10 \pm 0.02} & \mathbf{0.79 \pm 0.02} \end{bmatrix}$ | $\mathbf{0.11 \pm 0.02}$ | $\mathbf{1.59 \pm 0.01}$ |

**Table S3: Fitted coefficients polycrystalline $MAPbI_3$, illuminated small grain (Figure S11)**

| Diffusion Tensor [cm$^2$s$^{-1}$] | Diffusivity (coupled grain) [cm$^2$s$^{-1}$] | k$_1$ [$\times$ $\mathbf{10^6}$ s$^{-1}$] |
|---|---|---|
| $\begin{bmatrix} \mathbf{0.48 \pm 0.04} & \mathbf{0.15 \pm 0.04} \\ \mathbf{0.15 \pm 0.04} & \mathbf{0.83 \pm 0.04} \end{bmatrix}$ | $\mathbf{0.10 \pm 0.04}$ | $\mathbf{1.60 \pm 0.03}$ |

**Table S4: Diagonalized diffusion tensor and angle, illuminated small grain (Figure S11)**

| Diffusion Tensor [cm$^2$s$^{-1}$] | Angle between principal axes and image axes [°] |
|---|---|
| $\begin{bmatrix} \mathbf{0.89 \pm 0.06} & \mathbf{0} \\ \mathbf{0} & \mathbf{0.42 \pm 0.07} \end{bmatrix}$ | $\mathbf{69.1 \pm 3.3}$ |

**S8. Support Plane Analysis**

Custom Python code was written to perform support plane error analysis on the values of $P_{esc} \cdot k_2$ and $k_3$ and was used to determine the 95% confidence intervals using the methodology described in the PicoQuant Fluofit Manual (see section 6.3.1- Support Plane Method)[33]. Here, the tolerance of the reduced chi-squared value, $\chi^2_{tol}$ was determined using the equation:

$$\frac{\chi^2_{tol}}{\chi^2_{min}} = 1 + \frac{p}{v} F(p, v, P) \qquad \text{(S15)}$$

The confidence interval surfaces are determined by defining a tolerance over which the error becomes unacceptable. The tolerance level is determined from F-statistics, where $F(p, v, P)$ takes into account the number of parameters, $p$, the degrees of freedom, $v$, and a probability $P$ determined by the confidence interval of interest (i.e. 95%). For our simulation and fitting, $\nu = fitted\ pixels \times number\ of\ time\ points = 902 \times 3860$. Thus $F(5, 902 \times 3860, 0.95) = 0.7637$ and $\frac{\chi^2_{tol}}{\chi^2_{min}} = 1.0000015$.

**S9. Model Sensitivity to $P_{esc}$, $k_2$ and $k_3$**

Support plane analysis was performed as detailed in Section S9 by allowing the parameters $P_{esc} \times k_2$ and $k_3$ to vary. The resultant reduced $\chi^2$ surface as a ratio of the optimized reduced $\chi^2 = \chi^2_{opt}$ for the fit in Fig. 3 of the main text is shown in Fig. S14a). Clearly, there is no dependence on the value of $k_3$ used in the model up until $k_3 > 1 \times 10^{-28}$. However, we note that previous studies by Staub *et al.* have shown that the maximum possible value of $k_3 \approx 1 \times 10^{-28}$ given the trap densities and the trap depths reported in literature[20]. With a 95% confidence interval threshold and the upper limit for $k_3$, its value can range from $1 \times 10^{-32}$ $cm^6s^{-1}$ to $1 \times 10^{-28}$ $cm^6s^{-1}$ without impacting the resultant values of our model. Within this range, the 95% confidence interval of $P_{esc} \times k_2$ ranges from $1.84 \times 10^{-11}$ $cm^3s^{-1}$ to $4.14 \times 10^{-11}$ $cm^3s^{-1}$. If the $k_2$ value was fixed at $2 \times 10^{-10}$, this would mean the 95% CI for $P_{esc}$ would be 14.95 ± 5.75 %. On the other hand, if we use a $P_{esc} = 14\%$ based off the work of Pazos-Outón *et al*. and our previous work, as mentioned in Section S1, that would allow the range of $k_2$ to vary between $1.31 \times 10^{-10}$ $cm^3s^{-1}$ to $3.00 \times 10^{-10}$ $cm^3s^{-1}$ without impacting our fitted D and $k_1$ coefficients. These values lie well within the range of reported $k_2$ values in literature and is consistent with previously calculated values for our films through experimentally determined absorption coefficients and reciprocity as shown in our previous work[14,17].

We then perform fits to the data for different values of $P_{esc} \times k_2$ outside of the 95% CI range determined from the support plane analysis. The values chosen were $P_{esc} \times k_2 = 1.4 \times 10^{-11}$ $cm^3s^{-1}$ and $P_{esc} \times k_2 = 1.0 \times 10^{-10}$ $cm^3s^{-1}$. Due to the lack of sensitivity of the model to $k_3$, it was fixed at $1 \times 10^{-29}$. The results of the fits are summarized in table S5, with the fitted result in the main text shown in the first row. As evidenced from the fitted results, the diffusion tensor is still well constrained by the global fit to the data for all the pixels, even if the value of $P_{esc} \times k_2$ varies, given that the 2D map contains spatial information and their interdependence.

**Table S5: Fitted diffusion tensor for polycrystalline $MAPbI_3$, with varying values of $P_{esc} \times k_2$**

| $P_{esc} \times k_2$ [$cm^3s^{-1}$] | Diagonalized Diffusion Tensor [$cm^2s^{-1}$] | $k_1$ [$\times 10^6$ $s^{-1}$] | $\chi^2/\chi^2_{opt}$ |
|---|---|---|---|
| $2.8 \times 10^{-11}$ | $\begin{bmatrix} 0.93 & 0 \\ 0 & 0.72 \end{bmatrix}$ | 1.56 | 1 |
| $2.0 \times 10^{-12}$ | $\begin{bmatrix} 0.91 & 0 \\ 0 & 0.74 \end{bmatrix}$ | 2.22 | 1.0266 |
| $1.0 \times 10^{-10}$ | $\begin{bmatrix} 0.91 & 0 \\ 0 & 0.72 \end{bmatrix}$ | 0.13 | 1.2301 |

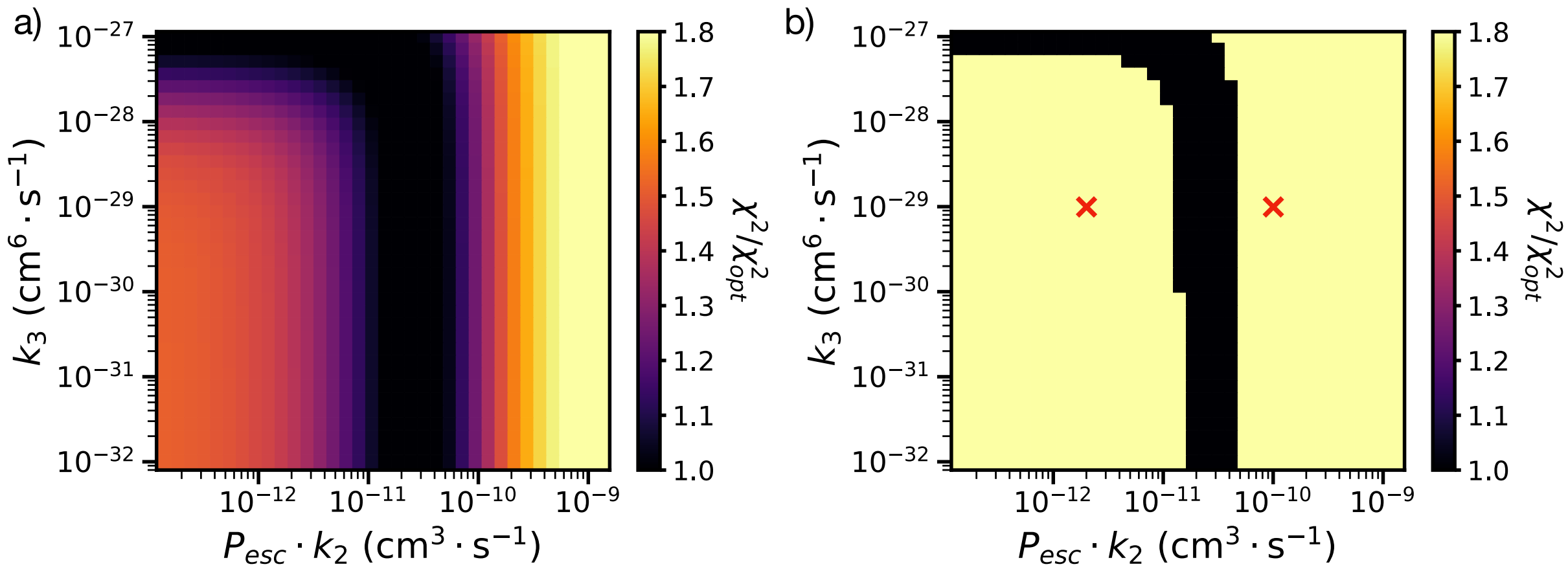

**Figure S15.** a) Reduced $\chi^2$ surface as a ratio of $\chi^2_{opt}$ where $\chi^2_{opt}$is the minimized $\chi^2$ value from the non-linear least squares fit as the free variables $P_{esc} \times k_2$ and $k_3$ are varied. We have multiplied together $P_{esc}$ and $k_2$ since their product is the final parameter input into Eq. S1. b) Support plane analysis of the $P_{esc} \times k_2$ and $k_3$ parameters, where the values are thresholded at a confidence interval of 95%. The $P_{esc} \times k_2$ values chosen for the fits in table S5 are shown with red crosses on the heatmap. We note that the maximum possible value of $k_3 \approx 1 \times 10^{-28}$ for $MAPbI_3$ given the reported trap depths and densities[20]. With this limit in mind, the 95% confidence range for $P_{esc} \times k_2$ ranges from $1.84 \times 10^{-11}$ $cm^3s^{-1}$ to $4.14 \times 10^{-11}$ $cm^3s^{-1}$ and the 95% confidence range for $k_3$ along with its limit ranges from $1 \times 10^{-32}$ $cm^6s^{-1}$ to $1 \times 10^{-28}$ $cm^6s^{-1}$.

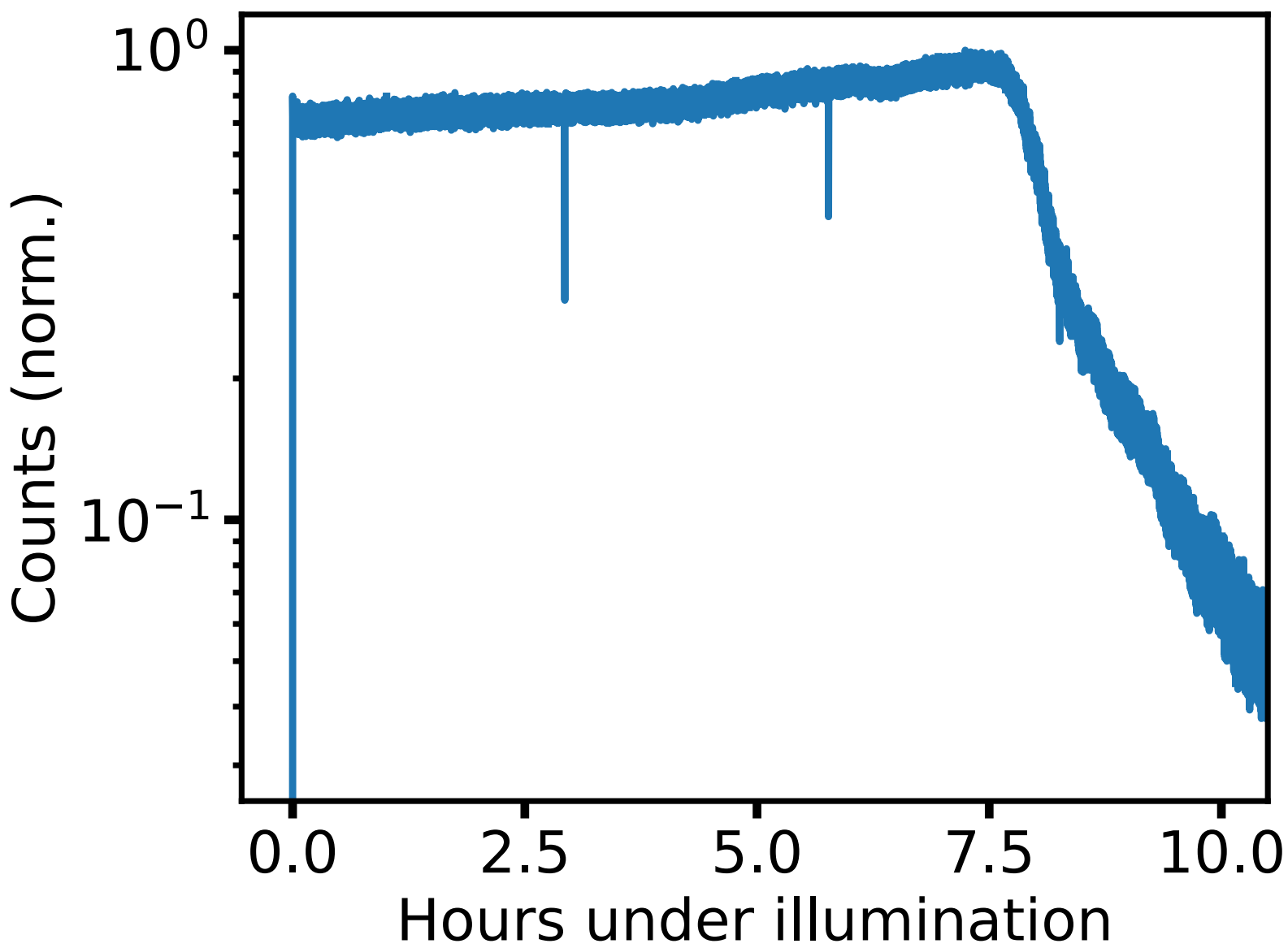

**Figure S16.** Photoluminescence (PL) counts of an unencapsulated $MAPbI_3$ thin-film under constant illumination for a carrier density ~$4x10^{17}cm^{-3}$ under ambient air conditions. Nearly constant PL with little photobrightening is observed until a drop-off after ~8 hours of constant illumination

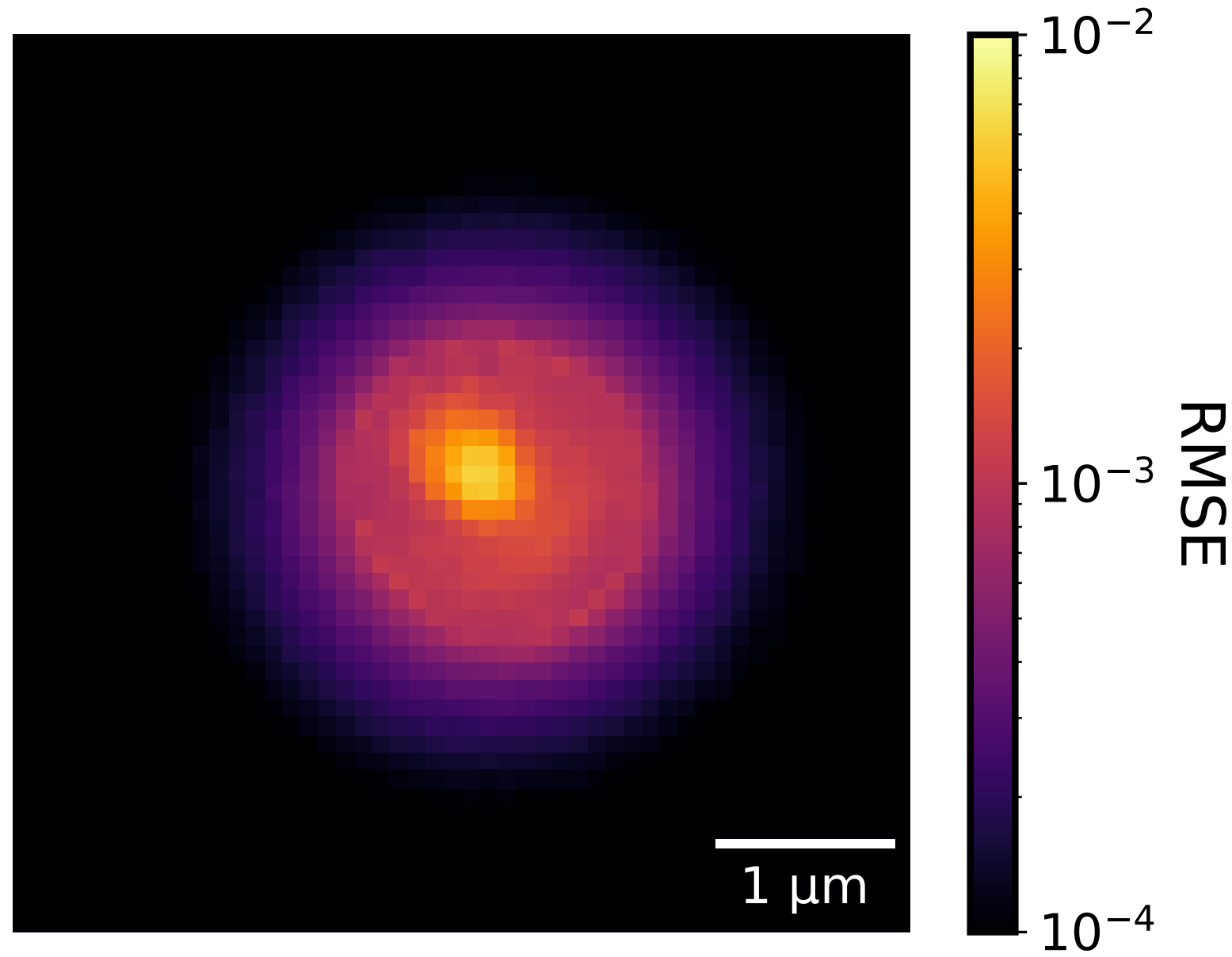

**Figure S17.** Summed root-mean-square-error as a function of position for the fitted diffusion data for the $MAPbBr_3$ single crystal shown in Fig. 2 in the main text. The total RMSE for the entire image is 0.0273.

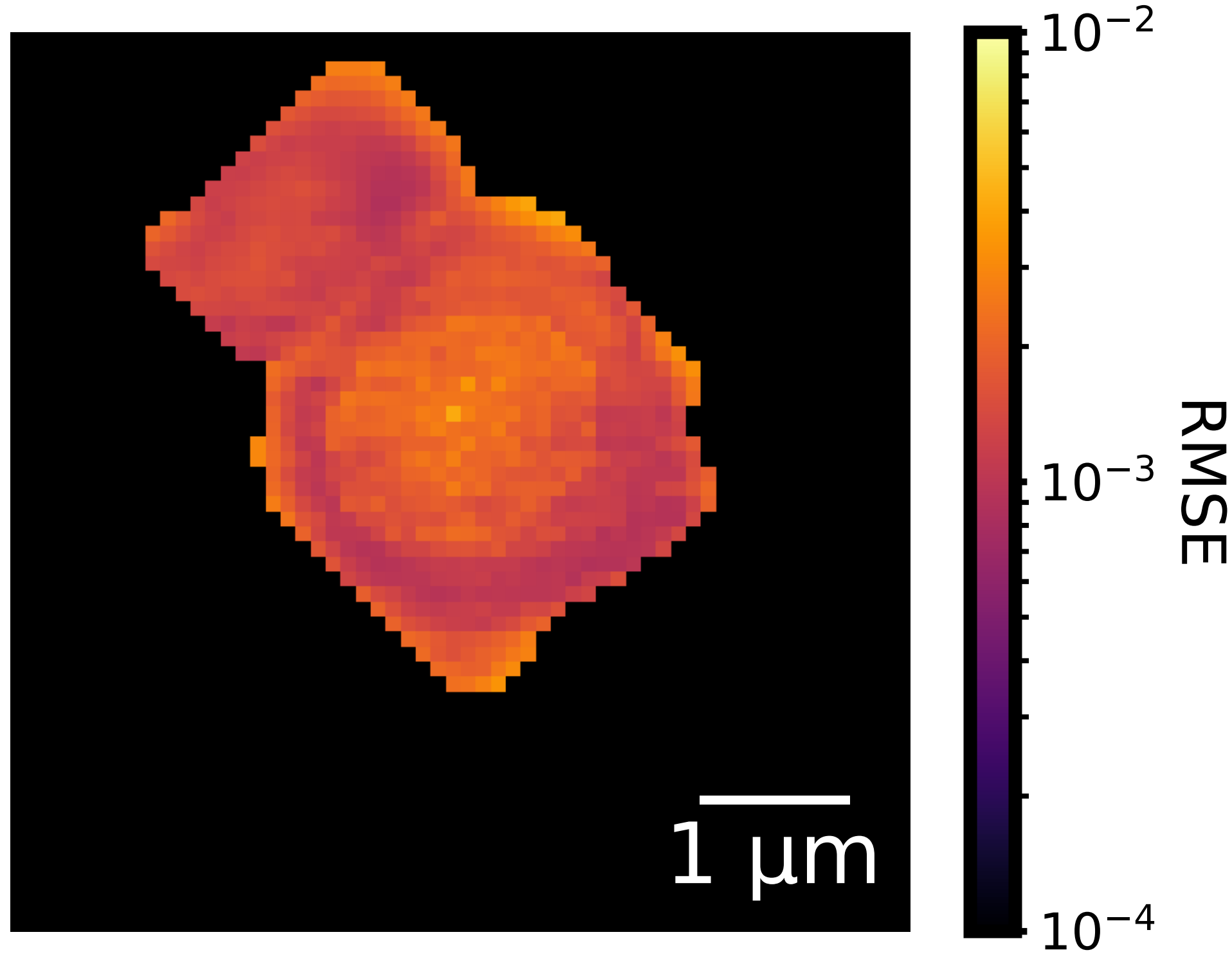

**Figure S18.** Temporally summed root-mean-square-error (RMSE) as a function of position for the fitted diffusion data for the $MAPbI_3$ polycrystalline thin-film shown in Figure 3 in the main text. The total RMSE for the entire image is 0.0511.

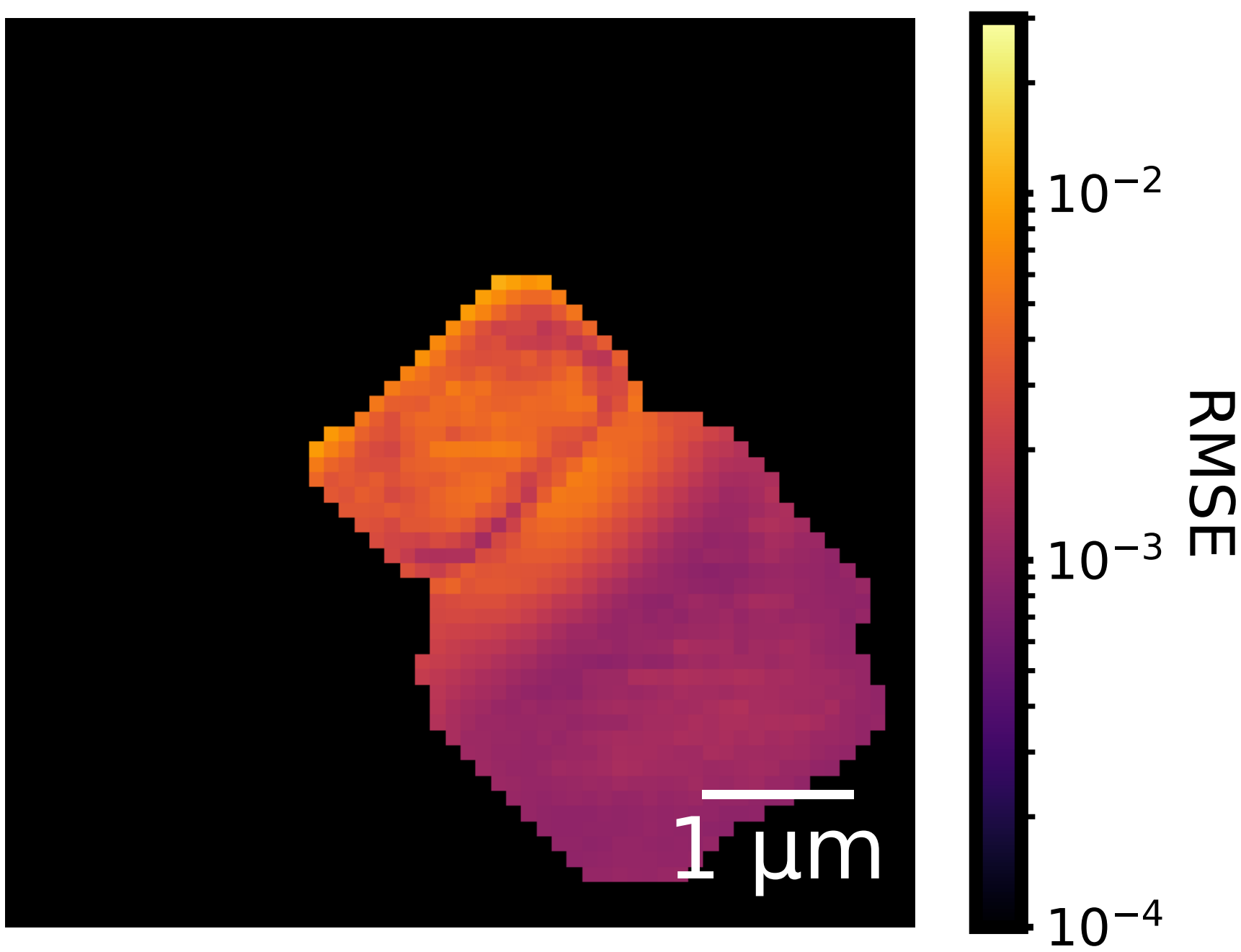

**Figure S19.** Temporally summed root-mean-square-error (RMSE) as a function of position for the fitted diffusion data for the $MAPbI_3$ polycrystalline thin-film shown in Fig. S13. The total RMSE for the entire image is 0.0804.

**Supporting Information References**